\documentclass[10pt]{article}
\usepackage[utf8]{inputenc}
\usepackage[T1]{fontenc}
\usepackage{arxiv}
\usepackage{mathptmx}

\usepackage{amsmath,amssymb}
\usepackage{graphicx}
\usepackage{booktabs,longtable,array}
\usepackage{microtype}
\usepackage{placeins}
\usepackage{caption}
\usepackage{cite}
\usepackage{xurl}
\usepackage[hidelinks]{hyperref}
\newcolumntype{L}[1]{>{\raggedright\arraybackslash}p{#1}}
\newcolumntype{C}[1]{>{\centering\arraybackslash}p{#1}}
\newcolumntype{R}[1]{>{\raggedleft\arraybackslash}p{#1}}
\newlength{\tablewidth}
\renewcommand{\headeright}{}
\renewcommand{\shorttitle}{\footnotesize Zeus-256: room-temperature NV-diamond quantum processor}
\renewcommand{\undertitle}{Preprint}
\AtBeginDocument{\raggedbottom}
\hypersetup{pdftitle={Experimental Benchmarking of a 256-Qubit Room-Temperature NV-Diamond Quantum Processor with an Optically Programmable Biomolecular Quantum Bus},pdfauthor={Adams Ivanov; Daniela Herrmann; Erick Giovani Sperandio Nascimento},pdfsubject={Zeus-256 NV-diamond processor manuscript}}
\title{Experimental Benchmarking of a 256-Qubit Room-Temperature NV-Diamond Quantum Processor with an Optically Programmable Biomolecular Quantum Bus}
\author{Adams Ivanov\textsuperscript{1}, Daniela Herrmann\textsuperscript{1},\\ Erick Giovani Sperandio Nascimento\textsuperscript{2,3}\\[5pt]
\normalfont\small\textsuperscript{1}Dynex Quantum Technologies AG, Switzerland\\
\normalfont\small\textsuperscript{2}Surrey Institute for People-Centred Artificial Intelligence,\\
\normalfont\small Faculty of Engineering and Physical Sciences, University of Surrey,\\
\normalfont\small Guildford GU2 7XH, United Kingdom.\\
\normalfont\small\textsuperscript{3}SENAI CIMATEC University, Salvador, BA, Brazil.\\
\normalfont\small Corresponding author: \href{mailto:office@dynex.ch}{office@dynex.ch}}
\date{11 October 2026}
\makeatletter
\renewcommand{\@maketitle}{%
  \begin{center}
  {\LARGE\bfseries\@title\par}\vspace{10pt}
  {\normalsize\begin{tabular}{c}\@author\end{tabular}\par}\vspace{7pt}
  {\small Preprint --- \@date\par}
  \end{center}\vspace{4pt}}
\makeatother
\begin{document}
\maketitle
\begin{abstract}
We report Zeus-256, a room-temperature nitrogen-vacancy (NV) diamond processor with 256 simultaneously active processor-level logical qubits arranged in a 16 x 16 array. Each logical qubit combines an NV electron-spin ensemble with nuclear-spin memory and ancillary resources. OpenQASM 2.0 circuits are transpiled into RFSoC-controlled microwave, radio-frequency, optical, timing, measurement and bus-configuration sequences. An optically programmable proprietary protein-chromophore layer with an internal metastable-triplet spin mediator provides a reconfigurable biomolecular quantum bus for local and long-range interactions. Benchmarks cover QFT, GHZ entanglement, Grover search, arithmetic, variational, neural-network, graph-state and Shor-family circuits. Factor recovery is reported for 11 integers of 8-84 bits, with 10 attempts per instance; the 84-bit endpoint uses 256 qubits and 40,250,342 reported gates, with a reported execution time of 551.694 ms. Bell-CHSH measurements yield \(S = 2.8008 \pm 0.0446\). Three-qubit Grover search returns the marked state in 958 of 1,024 shots (93.55\%), compared with an ideal reference of 94.53\%. A separate 30,000-shot population-and-parity protocol yields fixed-phase GHZ+ fidelity of 99.8960\%, an estimated standard error of 0.2885 percentage points and a conservative physical 95\% interval of [98.2342\%, 100\%]. Timing metrics are partial and do not establish quantum advantage. Scaling to 9,984 processor-level logical qubits is an engineering target, distinct from the demonstrated register.
\end{abstract}


\newpage

\FloatBarrier

\section{Introduction}\label{sec:1}

\FloatBarrier

\subsection{Problem statement and motivation}\label{sec:1.1}

Scalable quantum processors must simultaneously address qubit coherence, control fidelity, readout, inter-qubit connectivity, routing overhead, and the systems complexity required to coordinate large registers. In fixed-connectivity architectures, nonlocal interactions must be synthesized through routing operations, which can increase circuit depth and accumulated error \cite{ref9}. NV centres in diamond offer an alternative solid-state platform in which electron spins can be optically initialized, measured, and controlled with microwave fields, while associated nuclear spins provide longer-lived memory and ancillary degrees of freedom. Dutt et al. demonstrated an electron--nuclear NV quantum register \cite{ref1}, Neumann et al. demonstrated coherent control of coupled electron spins at room temperature \cite{ref2}, Dolde et al. demonstrated room-temperature entanglement between defect spins \cite{ref3}, and Blok et al. demonstrated measurement-based manipulation with real-time feedback \cite{ref4}. The remaining challenge is to integrate these capabilities into a larger processor architecture with programmable connectivity while preserving room-temperature operation.

The experimental system investigated in this work, Zeus-256, is a room-temperature NV-diamond quantum processor comprising a 16 \(\times\) 16 array of 256 simultaneously active processor fields. Each field implements one processor-level logical qubit using an ensemble of NV electron spins as the computational degree of freedom and associated nuclear-spin degrees of freedom as memory and ancillary resources. A proprietary electron--nuclear-spin control protocol, including error-management functions, is applied at the logical-qubit control layer. The processor does not obtain its 256-qubit register through time multiplexing or qubit reuse.

A central feature of Zeus-256 is its optically programmable biomolecular quantum-bus coupling fabric. The biomolecular material is not treated merely as a passive dielectric: it uses a proprietary protein--chromophore architecture similar to EYFP and incorporates an optically addressable metastable electronic triplet spin degree of freedom as an internal molecular mediator. Fluorescent-protein chromophores such as EYFP have independently been demonstrated as optically addressable triplet-state spin qubits, including optical initialization/readout and coherent microwave control, establishing the physical plausibility of this class of biomolecular spin degree of freedom \cite{ref25}. In Zeus-256, the molecular spin degree of freedom is not exposed as a user-addressable processor qubit; instead, it forms part of the biomolecular quantum-bus subsystem. Optical programming changes the molecular material's dielectric and coupling state and thereby reshapes the microwave and radio-frequency electromagnetic environment across selected processor regions. The programmed molecular state can establish a shared interaction channel between selected processor fields, including spatially separated and non-nearest-neighbour fields, allowing quantum information to be exchanged through the bus and enabling mediated entangling operations. OpenQASM 2.0 circuits are transpiled into processor-specific RFSoC-controlled microwave, radio-frequency, optical, timing, measurement, and bus-configuration sequences, with individually calibrated control parameters applied to each processor field.

The experiments reported here are performed on the 256-qubit Zeus-256 implementation. A larger processor architecture targeting up to 9,984 processor-level logical qubits is discussed as an engineering scale-up architecture and is distinguished throughout from the experimentally demonstrated 256-qubit register.

\FloatBarrier

\subsection{Research questions and hypotheses}\label{sec:1.2}

This work addresses three research questions. RQ1 asks whether the NV-diamond electron--nuclear processor-level logical-qubit architecture can operate a 256-qubit register at room temperature and execute heterogeneous quantum circuits across the full register. RQ2 asks whether the integrated Zeus-256 architecture, incorporating its proprietary biomolecular quantum-bus coupling fabric, can execute compiled multi-qubit circuits requiring configurable interactions between both local and spatially separated processor fields across the demonstrated register. RQ3 asks how the recorded hardware execution wall time and fluorescence-data interpretation time vary with circuit width, gate count, compilation, and circuit structure for representative quantum workloads.

The corresponding hypotheses are: H1, that the processor can execute quantum circuits using up to 256 simultaneously active processor-level logical qubits at room temperature; H2, that the integrated processor can execute compiled multi-qubit circuits requiring configurable local and long-range interaction patterns through the proprietary biomolecular quantum bus and control stack; and H3, that hardware execution wall time is expected to vary jointly with circuit width, gate count, compilation, and circuit structure rather than with qubit count alone. The timing analysis is descriptive and is not formulated as a fitted scaling law.

\FloatBarrier

\subsection{Objectives}\label{sec:1.3}

The overall objective is to experimentally characterize Zeus-256 as a room-temperature, programmable-connectivity NV-diamond quantum processor. The specific objectives are to characterize the underlying NV-spin and processor-level logical-qubit control system; demonstrate operation of a 256-qubit register; evaluate the integrated proprietary biomolecular quantum-bus and control stack at the observable processor boundary; describe the OpenQASM-to-pulse execution pathway; benchmark representative circuit families at increasing width and depth; quantify hardware execution wall time and fluorescence-data interpretation time under the timing definitions used in this work; evaluate output-state quality using appropriate observables and fidelity measures; and identify the present limitations and scaling requirements of the architecture.

\FloatBarrier

\subsection{Contributions and novelty}\label{sec:1.4}

The principal contribution of this work is the experimental implementation and characterization of a 256-qubit room-temperature NV-diamond processor that combines ensemble-based electron-spin computation, nuclear-spin memory and ancillary degrees of freedom, MW/RF pulse control, optical initialization and fluorescence readout, proprietary electron--nuclear-spin logical control, and an optically programmable biomolecular quantum-bus coupling fabric within a single processor architecture. In contrast to fixed-connectivity architectures, where hardware graph constraints can impose routing overhead \cite{ref9}, and using a different physical mechanism from reconfigurable neutral-atom and trapped-ion systems \cite{ref10,ref11}, Zeus-256 uses the programmed state of the biomolecular material to provide an effective quantum bus. This shared coupling channel enables selected processor fields, including non-nearest-neighbour fields, to exchange quantum information and participate in mediated entangling operations while the optical state of the material determines which effective interaction pathways are made available to the compiled circuit.

A second contribution is the experimental benchmarking of the processor across heterogeneous completed circuit classes, including quantum Fourier transform, multipartite entanglement, Grover search, arithmetic and miscellaneous QASM circuits, graph-state preparation, variational quantum eigensolver circuits, quantum neural network circuits, portfolio-VQE, travelling-salesman, and routing circuits. The benchmark suite includes circuits extending to 256 processor-level logical qubits and is used to analyze processor execution and fluorescence-data interpretation behavior across differing circuit structures. Section~\ref{sec:6.6} reports the completed Shor-family factorization staircase, using a Gidney--Ekerå-inspired Ekerå--Håstad implementation with windowed exact arithmetic. Factor recovery is reported for all 11 test integers from 8 to 84 bits, with register widths extending from 28 to all 256 processor-level logical qubits. A third contribution is the separation of experimentally demonstrated processor scale from architectural scaling projections: all completed processor-level experiments reported in this work are attributed explicitly to Zeus-256, while larger register sizes are treated as architectural targets rather than as demonstrated hardware.

\FloatBarrier

\section{State of the Art and Research Gaps}\label{sec:2}

\FloatBarrier

\subsection{NV centres as electron--nuclear quantum registers}\label{sec:2.1}

NV centres provide an experimentally established route to combining optically addressable electron spins with longer-lived nuclear-spin degrees of freedom. Dutt et al. demonstrated a quantum register in which the electron spin of an NV centre was coherently controlled together with nearby nuclear spins, establishing the electron--nuclear register model that underlies many subsequent diamond-spin architectures \cite{ref1}. Neumann et al. subsequently demonstrated coherent control of coupled electron spins in a room-temperature solid-state register \cite{ref2}. Dolde et al. experimentally generated entanglement between two engineered NV electron-spin qubits under ambient conditions and verified the state by quantum-state tomography; the reported entanglement fidelity was 0.67 \(\pm\) 0.04, and entanglement storage was extended by transferring the state to nuclear-spin degrees of freedom \cite{ref3}.

Work on larger local spin registers has progressively increased the number of controllable degrees of freedom. Taminiau et al. used an NV electron spin to initialize, control, and read out weakly coupled carbon-13 nuclear spins, implemented high-fidelity single- and two-qubit gates, and demonstrated a three-qubit quantum-error-correction protocol \cite{ref5}. Bradley et al. later realized a 10-qubit solid-state register consisting of one NV electron spin and nine nuclear spins, demonstrated pairwise connectivity across all 45 qubit pairs, generated genuine multipartite entanglement involving up to seven qubits, and reported nuclear-spin quantum-memory coherence extending to 63(2) s \cite{ref6}. These studies establish the feasibility of coherent electron--nuclear registers at room temperature while also illustrating the experimental control and connectivity challenges that arise as register size increases.

\FloatBarrier

\subsection{Error correction and logical qubits in diamond}\label{sec:2.2}

Quantum error correction has been demonstrated experimentally in NV-diamond spin registers, but the term logical qubit requires a precise operational definition. The three-qubit protocol of Taminiau et al. encoded information redundantly across nuclear-spin degrees of freedom and demonstrated increased robustness against applied errors \cite{ref5}. Abobeih et al. subsequently implemented fault-tolerant operations on a logical qubit in a diamond quantum processor using a five-qubit code with a flag protocol requiring seven physical spin qubits in total \cite{ref7}. Their experiment included repeated multi-qubit measurements, fault-tolerant logical Clifford operations, flagged stabilizer measurements, and real-time processing, while also noting that further fidelity improvements were required for logical error rates to fall below the physical error rates \cite{ref7}.

These results provide an important reference point for the terminology used in the present work. In Zeus-256, each processor field exposes one processor-level logical qubit formed from an ensemble of NV electron spins used as the computational degree of freedom together with associated nuclear-spin memory and ancillary resources. The proprietary logical-qubit control protocol used by Zeus-256 is therefore treated as part of the processor implementation and must be characterized experimentally rather than inferred solely from the use of the term logical qubit. Section~\ref{sec:5} describes the corresponding construction, control, and validation methodology.

\FloatBarrier

\subsection{Measurement-conditioned and dissipative quantum control}\label{sec:2.3}

Controlled dissipation and measurement feedback provide an alternative to viewing all non-unitary dynamics solely as unwanted decoherence. Verstraete, Wolf, and Cirac established a theoretical framework in which engineered dissipation can be used for quantum-state preparation and computation by driving a system toward selected steady states \cite{ref8}. In an NV-diamond experiment, Blok et al. demonstrated adaptive manipulation of a nuclear-spin qubit through sequential partial measurements and real-time feedback, showing that measurement backaction can itself function as a controlled resource \cite{ref4}.

The Zeus architecture builds on this broader control paradigm by combining coherent microwave and radio-frequency excitation with optical initialization, fluorescence readout, measurement-conditioned control, and programmable inter-field coupling. The distinction between coherent pulse operations, measurement-conditioned feedback, and the optically programmed biomolecular quantum-bus coupling fabric is retained explicitly in the experimental methodology so that these mechanisms are not conflated in the interpretation of integrated processor-level measurements.

\FloatBarrier

\subsection{Connectivity, routing overhead, and reconfigurable interaction architectures}\label{sec:2.4}

Connectivity is a fundamental system-level constraint in quantum processors because an algorithmic two-qubit interaction can be executed directly only when the hardware provides an appropriate physical or effective interaction pathway between the participating qubits. When the interaction graph required by a quantum circuit differs from the connectivity graph exposed by the processor, compilation generally introduces routing operations such as SWAP gates, increasing circuit depth, execution time, and accumulated error. Yuan and Zhang formally characterized this depth overhead for quantum-circuit compilation under arbitrary connectivity constraints, demonstrating that limitations of the underlying interaction graph can impose substantial and, in some cases, asymptotically significant routing overhead \cite{ref9}.

Several quantum-computing modalities address this limitation by providing dynamically reconfigurable or highly connected interaction architectures. Bluvstein et al. demonstrated a logical processor based on reconfigurable neutral-atom arrays in which atomic rearrangement enables programmable connectivity, with experiments involving up to 48 logical qubits \cite{ref10}. Ransford et al. reported a 98-qubit trapped-ion quantum computer providing effectively all-to-all connectivity through ion transport within a quantum charge-coupled-device architecture \cite{ref11}. These systems illustrate an important architectural principle: physical or effective reconfiguration of the interaction graph can reduce the requirement to synthesize nonlocal circuit interactions through long sequences of nearest-neighbour routing operations.

The Zeus-256 architecture addresses the same processor-level connectivity problem using a fundamentally different mechanism appropriate to a solid-state NV-diamond system. Rather than physically transporting qubits, the processor incorporates an optically programmable biomolecular quantum bus spanning the processor-field array. The proprietary protein--chromophore coupling layer contains an optically addressable metastable electronic triplet-spin degree of freedom that functions internally as a molecular mediator, while optical programming also modifies the dielectric and electromagnetic coupling environment. Together, these mechanisms establish a shared, reconfigurable interaction resource through which selected processor-level logical-qubit fields, including spatially separated and non-nearest-neighbour fields, can participate in mediated interactions.

At the compiler and control level, the available bus-mediated pathways are represented through a calibrated effective interaction map rather than a fixed nearest-neighbour lattice. Circuit interactions can therefore be mapped onto local or long-range bus configurations together with the corresponding MW/RF and optical control sequences. This does not imply unrestricted instantaneous all-to-all coupling, nor does it disclose a particular microscopic molecule-to-spin Hamiltonian; rather, it defines the processor-level connectivity abstraction used by Zeus-256. The relevant distinction from neutral-atom and trapped-ion approaches is therefore not the objective of reconfigurable connectivity, but the physical mechanism by which that connectivity is realized: Zeus uses an optically programmed biomolecular spin mediator together with an optically reconfigured dielectric/electromagnetic environment within a stationary solid-state NV-diamond register.

\FloatBarrier

\subsection{Dielectric engineering and biomolecular spin mediators}\label{sec:2.5}

The electromagnetic environment surrounding an NV centre strongly influences the spatial distribution, strength, and homogeneity of the microwave fields used for spin control. Le Floch et al. demonstrated single-NV optically detected magnetic resonance using a tunable dielectric microwave cavity and experimentally mapped the position dependence of the cavity magnetic field \cite{ref12}. More recently, Hermann et al. used a dielectric resonator to generate strong and spatially homogeneous microwave fields over an NV ensemble, reaching Rabi frequencies of approximately 85 MHz \cite{ref13}. These studies establish dielectric structures as an experimentally relevant tool for engineering microwave delivery to NV spins.

Fluorescent-protein chromophores provide a separate but complementary precedent for biomolecular quantum degrees of freedom. Feder et al. demonstrated an optically addressable spin qubit in enhanced yellow fluorescent protein (EYFP), using the chromophore's metastable electronic triplet state for spin initialization, readout, and coherent microwave manipulation; the same work also demonstrated room-temperature optically detected magnetic resonance \cite{ref25}. The Zeus coupling material is not identified as EYFP. Rather, it employs a proprietary protein--chromophore system similar to this class of fluorescent-protein spin qubits, with a metastable triplet spin degree of freedom used internally as part of the biomolecular quantum-bus subsystem.

The coupling mechanism used in Zeus-256 differs from static or mechanically tuned dielectric structures. The proprietary biomolecular material uses a protein--chromophore architecture similar to EYFP and combines two processor-level functions: optically programmable dielectric reshaping of the MW/RF environment and an internal metastable-triplet spin degree of freedom that participates in the biomolecular quantum-bus subsystem. We refer to the resulting shared, reconfigurable interaction channel as a biomolecular quantum bus: it permits mediated exchange of quantum information between spatially separated processor-level logical qubits and provides the interaction resource used by compiled multi-qubit and entangling operations. The term quantum bus is used here at the processor-functional level and does not imply disclosure of the exact molecular identity, microscopic triplet-state dynamics, propagating mode, bus-transfer Hamiltonian, or molecule-to-spin coupling mechanism. The EYFP result in Ref. \cite{ref25} is cited as an external precedent for an optically addressable fluorescent-protein triplet spin qubit, not as an identification of the proprietary Zeus material. The exact molecular composition, microscopic bus dynamics, calibrated transfer characteristics, programming procedure, and native entangling implementation remain proprietary.

\FloatBarrier

\subsection{Research gaps and positioning of this work}\label{sec:2.6}

The literature above identifies four gaps relevant to the present study. First, landmark room-temperature NV-diamond experiments have demonstrated coherent electron--nuclear registers, multipartite entanglement, quantum error correction, and fault-tolerant logical operations, but experimentally controlling substantially larger simultaneously active logical registers remains a systems challenge \cite{ref1,ref2,ref3,ref5,ref6,ref7}. Second, dynamic connectivity has been demonstrated at larger scale in neutral-atom and trapped-ion processors, whereas processor-level long-range coupling between spatially separated fields in a solid-state NV-diamond register remains comparatively unexplored \cite{ref10,ref11}. Third, dielectric structures are established for shaping and homogenizing microwave fields in NV systems \cite{ref12,ref13}, and fluorescent-protein chromophores have now been demonstrated as optically addressable metastable-triplet spin qubits \cite{ref25}, but integration of these concepts into an optically programmable biomolecular spin mediator that functions as a shared quantum-information bus within a compiled NV-diamond processor remains an open systems problem. Fourth, a large-register architecture must be evaluated not only by nominal qubit count but also by state quality, circuit execution, readout, stability, long-range interaction capability, and reproducible processor-level benchmark methodology.

Zeus-256 is positioned against these gaps as an experimental 256-logical-qubit room-temperature NV-diamond processor. The present work therefore focuses on five forms of evidence: simultaneous operation of the 256-field register; characterization of the electron--nuclear logical-qubit implementation; processor-boundary evaluation of an integrated proprietary biomolecular quantum-bus and control stack capable of addressing spatially separated processor fields; entanglement and output-state validation; and circuit-level benchmarking across heterogeneous workloads. Larger register sizes discussed later in the paper are treated as architectural scaling targets and are not used as substitutes for the experimentally demonstrated Zeus-256 results.

\FloatBarrier

\section{Processor Architecture}\label{sec:3}

\FloatBarrier

\subsection{Zeus-256 experimental device}\label{sec:3.1}

Zeus-256 is the experimentally implemented processor studied in this work. The device is built around a 10 mm \(\times\) 10 mm \(\times\) 0.5 mm CVD diamond containing an active population of nitrogen-vacancy centres and is operated at room temperature under a static bias magnetic field. The active processor plane is organized as a 16 \(\times\) 16 array of 256 independently controlled processor fields. Each field corresponds to one processor-level logical qubit, and all 256 processor-level logical qubits are physically active simultaneously; the processor therefore does not obtain its qubit count through time multiplexing, sequential reuse, or virtual expansion of a smaller register.

The processor integrates microwave control of NV electron-spin degrees of freedom, radio-frequency control of associated nuclear-spin degrees of freedom, optical initialization, spin-dependent fluorescence readout, an optically programmable biomolecular quantum-bus coupling layer with an internal protein--chromophore triplet-spin mediator, RFSoC-based pulse generation and orchestration, and measurement-conditioned feedback. The biomolecular subsystem is an internal interaction resource and is not counted as an additional processor-level logical-qubit register. The experimentally demonstrated 256-qubit device is distinguished throughout this paper from the 9,984-qubit engineering scale target discussed in Section~\ref{sec:3.8}.

\begin{figure}[htbp]
\centering
\renewcommand{\thefigure}{1}\renewcommand{\theHfigure}{1}
\includegraphics[width=\linewidth,height=0.69\textheight,keepaspectratio,trim={0.0000bp 118.0000bp 0.0000bp 0.0000bp},clip]{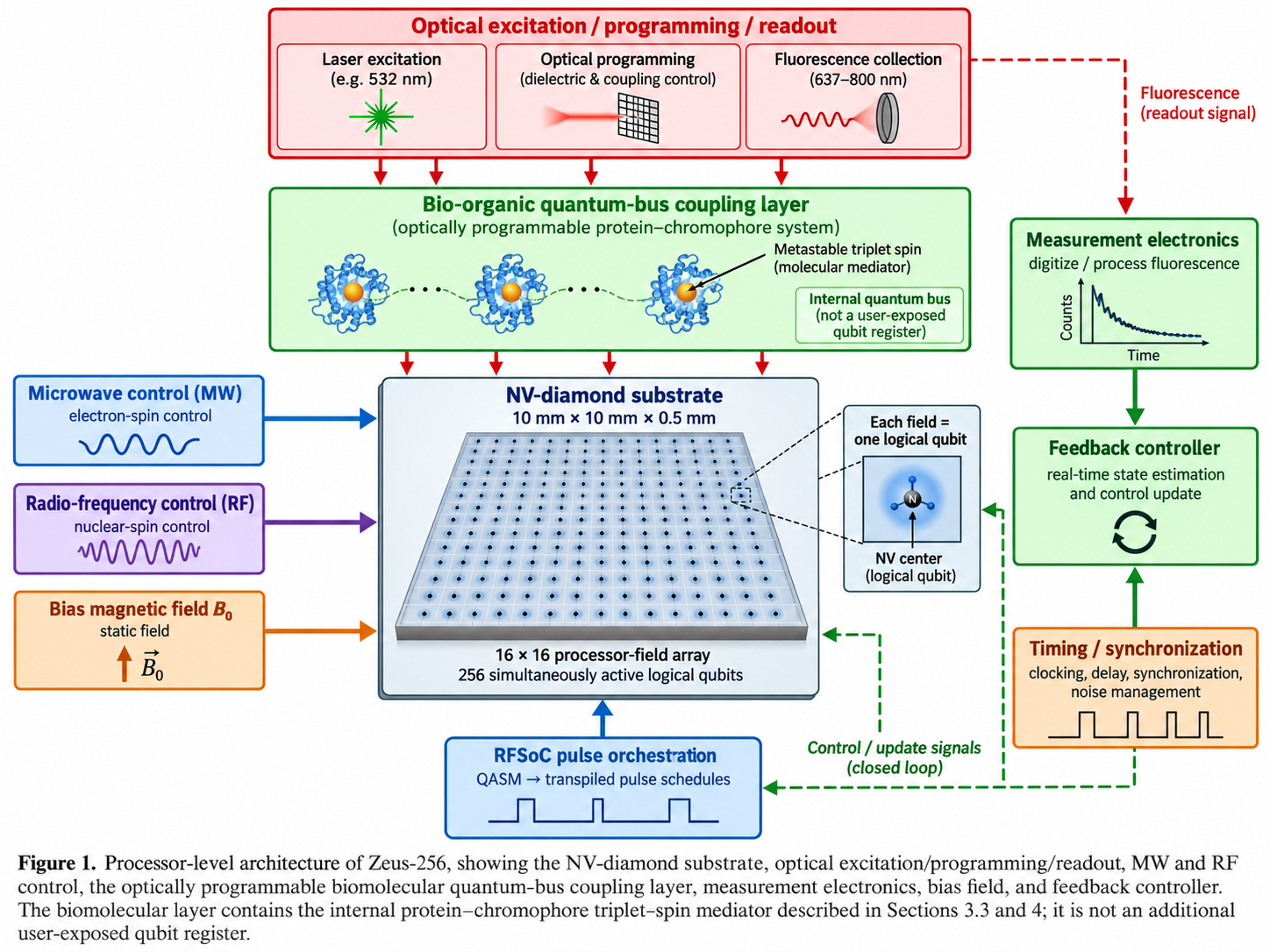}
\caption{Processor-level architecture of Zeus-256, showing the NV-diamond substrate, optical excitation/programming/readout, MW and RF control, the optically programmable biomolecular quantum-bus coupling layer, measurement electronics, bias field, timing/synchronization, RFSoC pulse orchestration, and the feedback controller. The biomolecular layer contains the internal protein--chromophore triplet-spin mediator described in Sections~\ref{sec:3.3} and~\ref{sec:4}; it is an interaction resource and not an additional user-exposed qubit register.}\label{fig:1}
\end{figure}

\FloatBarrier

\subsection{16 \(\times\) 16 logical-qubit array}\label{sec:3.2}

The diamond is partitioned at the processor-control level into 256 fields arranged in a 16 \(\times\) 16 grid. Each field is independently calibrated and receives its own microwave and radio-frequency control parameters. The field abstraction defines the addressable unit presented to the compiler and provides the physical organization through which local control, coupling configuration, initialization, and readout are coordinated.

Unlike a single-defect architecture in which one isolated NV centre is equated directly with one processor qubit, each Zeus-256 field contains an ensemble of NV-related spin degrees of freedom. The ensemble provides the computational spin population used by the logical-qubit control layer and increases the available optical signal for fluorescence-based state discrimination. Associated nuclear-spin degrees of freedom provide memory and ancillary resources \cite{ref1,ref5,ref6}. The resulting hierarchy is therefore: microscopic NV electron and nuclear spins, an ensemble-controlled processor field, and one processor-level logical qubit exposed to the quantum-circuit layer.

\begin{figure}[htbp]
\centering
\renewcommand{\thefigure}{2}\renewcommand{\theHfigure}{2}
\includegraphics[width=\linewidth,height=0.69\textheight,keepaspectratio,trim={0.0000bp 285.0000bp 0.0000bp 0.0000bp},clip]{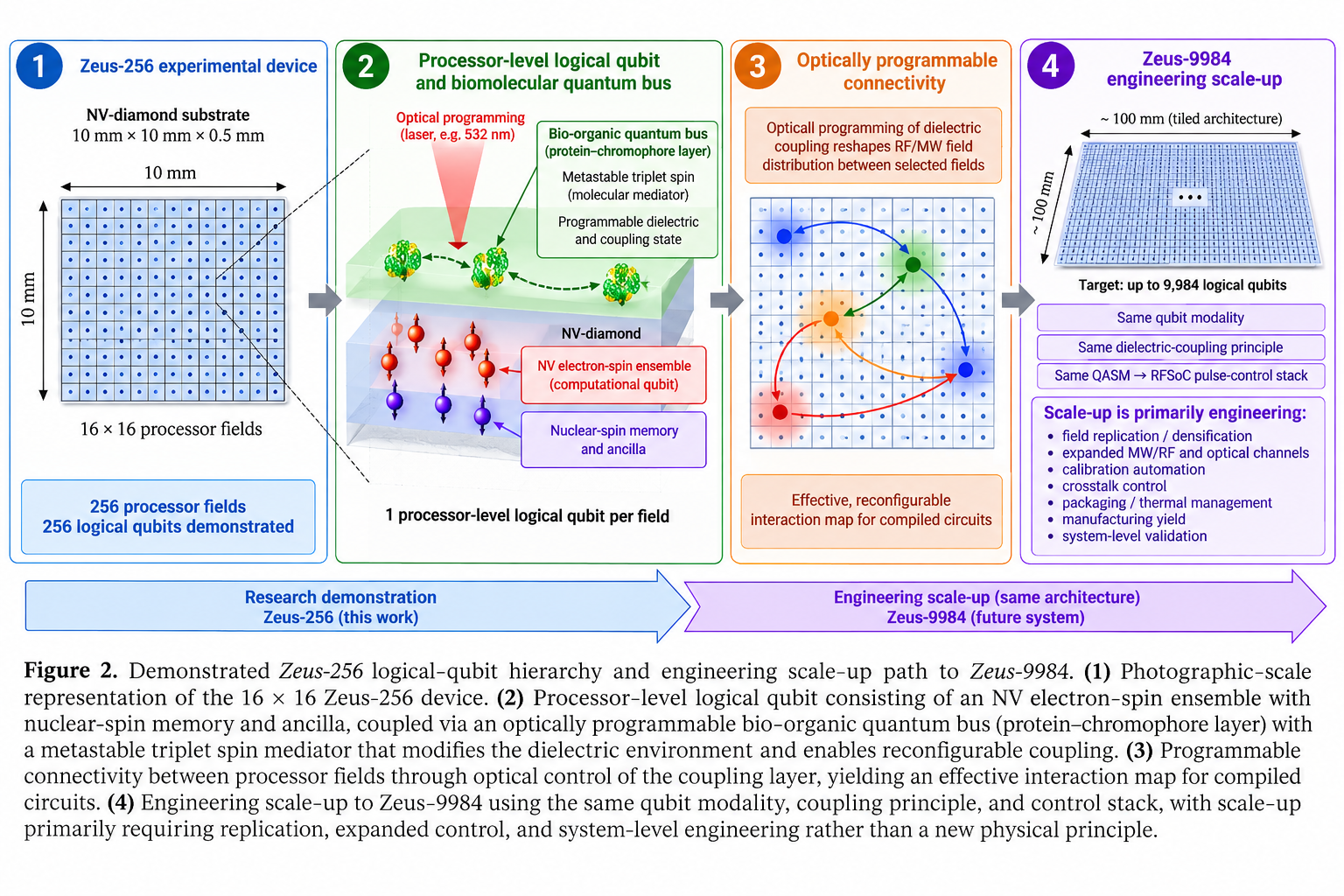}
\caption{Demonstrated Zeus-256 logical-qubit hierarchy and engineering scale-up path to Zeus-9984. The demonstrated 16 \(\times\) 16 Zeus-256 register and one processor-level logical-qubit field are shown together with the optically programmable biomolecular quantum-bus coupling layer, which contains an internal metastable electronic triplet-spin mediator and provides effective local and long-range interaction pathways for compiled circuits. The figure also shows the principal engineering tasks associated with scaling the same architecture toward the target of 9,984 processor-level logical qubits.}\label{fig:2}
\end{figure}

\FloatBarrier

\subsection{Optically programmable biomolecular quantum-bus coupling fabric}\label{sec:3.3}

Inter-field connectivity is implemented through a proprietary optically programmable biomolecular quantum-bus coupling layer integrated into the processor stack. The material uses a protein--chromophore architecture similar to EYFP rather than being disclosed as EYFP itself. Its metastable electronic triplet provides an optically addressable molecular spin degree of freedom that is used internally as part of the coupling fabric. In Zeus-256 this molecular spin degree of freedom is not counted among, or exposed as, the 256 processor-level logical qubits; it acts as an internal interaction mediator. Optical programming changes both the dielectric response and the effective coupling state of the biomolecular layer, reshaping the microwave and radio-frequency electromagnetic environment across the NV-diamond substrate. At the processor level, the programmed material functions as an effective quantum bus: a shared coupling resource that can connect selected processor fields without requiring those fields to be geometrically adjacent. The bus therefore supports mediated exchange of quantum information between spatially separated processor-level logical qubits and provides a mechanism for establishing the nonlocal interactions required by compiled multi-qubit gates and entanglement circuits.

At the processor level, the programmed biomolecular-bus state is represented internally by a proprietary calibrated effective interaction map C. An element \(C_{ij}\) denotes the internally calibrated bus-mediated interaction or control relationship available between processor-level logical-qubit fields i and j under a given optical programming configuration. Entries in C are not restricted to nearest-neighbour fields: selected spatially separated fields may be connected through the same biomolecular-bus layer. C is an internal processor abstraction rather than an assertion of a specific microscopic molecule-to-spin Hamiltonian, and the underlying calibration data are not disclosed in this work. Circuit compilation maps required local and nonlocal interactions onto proprietary optical bus-programming patterns together with the corresponding field-specific MW/RF pulse schedules.

The biomolecular quantum-bus coupling layer is distinct from measurement-conditioned feedback. Optical programming configures the molecular spin mediator together with the dielectric/electromagnetic environment and therefore the effective inter-field interaction pathways, whereas the feedback controller adapts subsequent local control parameters in response to measured spin states. The two mechanisms can operate jointly, allowing a programmed bus-mediated coupling topology to coexist with state-dependent microwave, RF, optical, timing, and noise updates.

\FloatBarrier

\subsection{OpenQASM-to-pulse control and RFSoC orchestration}\label{sec:3.4}

Quantum circuits are supplied to Zeus-256 in OpenQASM 2.0 (hereafter QASM) form \cite{ref23} and are transpiled by the Zeus control stack into processor-specific control schedules rather than being executed directly as abstract gates. The transpiled schedule specifies the microwave, radio-frequency, optical, timing, measurement, and biomolecular-bus configuration operations required by the mapped circuit. An RFSoC-based control layer generates and coordinates the resulting hardware pulse and control sequences.

Each of the 256 processor fields receives individually calibrated microwave and radio-frequency frequencies, amplitudes, phases, and timing parameters appropriate to its local electron- and nuclear-spin transitions. Circuit gate counts reported in the benchmark section refer to non-measurement gates in the original QASM circuit; final measurement statements are excluded unless explicitly stated otherwise. These counts do not represent the number of physical MW/RF pulse operations after transpilation. One algorithmic gate may map to multiple hardware-level pulses, while processor-specific compilation and pulse fusion can also combine control operations.

\FloatBarrier

\subsection{Optical initialization, fluorescence readout, and state estimation}\label{sec:3.5}

Optical excitation initializes the NV electron-spin population, while spin-dependent fluorescence provides the principal measurement channel. Excitation and fluorescence collection are mapped onto the 16 \(\times\) 16 processor-field array so that logical-qubit states can be estimated from ensemble fluorescence signals. The optical measurement stack conditions and digitizes the fluorescence response and converts it into processor-level observables and logical-state estimates. Established NV-spin experiments provide the underlying basis for optical initialization, spin-dependent fluorescence readout, and electron--nuclear register control \cite{ref1,ref3,ref5}.

In the benchmark timing tables, the reported measurement-related quantity is fluorescence-data interpretation time, \(T_{\mathrm{interpretation}}\). It denotes processing of already acquired fluorescence data and excludes the physical photon-acquisition interval. Consequently, \(T_{\mathrm{interpretation}}\) is not a complete measurement latency or end-to-end wall-clock latency.

\FloatBarrier

\subsection{Processor-level electron--nuclear logical qubits}\label{sec:3.6}

Each Zeus-256 processor-level logical qubit uses an ensemble of NV electron spins as its primary computational degree of freedom together with associated nuclear-spin degrees of freedom that act as memory and ancillary resources, consistent with established NV electron--nuclear register architectures \cite{ref1,ref5,ref6}. A proprietary electron--nuclear-spin control protocol, including error-management functions, is applied at the processor-level logical-qubit layer. The processor-level logical-qubit abstraction therefore represents an actively controlled processor variable rather than a direct count of individual microscopic NV defects.

The detailed implementation of the control protocol is proprietary, but its processor-level function is experimentally testable. Logical-qubit validation is based on repeated preparation, controlled evolution, fluorescence-based measurement, and comparison of logical-state performance under the applied control protocol. The experimental methodology therefore distinguishes raw spin populations, processor fields, and processor-level logical qubits rather than using these quantities interchangeably.

In this paper, the term processor-level logical qubit refers specifically to the processor variable implemented in each Zeus-256 field. It is an implementation term and is not used here to imply demonstrated fault-tolerant logical error suppression. The experiments reported in this work use 256 such simultaneously active processor-level logical qubits. The 9,984-qubit configuration discussed later is a scale-up target of the same architecture and is not presented as an experimentally demonstrated register.

\FloatBarrier

\subsection{Measurement-conditioned control}\label{sec:3.7}

Fluorescence-derived state variables are processed by a real-time controller that can update subsequent microwave, RF, optical, timing, delay, and noise parameters. This measurement-conditioned control operates in addition to the biomolecular quantum-bus coupling layer. A generic feedback relation is

\[
h_i(t)=b_i(t)+\sum_j J_{ij}(t)\,m_j(t-\tau_{ij})+\eta_i(t),
\]

where \(h_i\) is an effective control command, \(b_i\) is a local bias, \(J_{ij}\) is a programmable feedback coefficient, \(m_j\) is a measured or estimated state variable, \(\tau_{ij}\) is an optional delay, and \(\eta_i\) is an optional deterministic or stochastic term. The resulting control command is translated into physical pulse and measurement parameters.

Measurement-conditioned feedback complements the optically programmed biomolecular quantum-bus configuration. The bus layer configures the internal molecular mediator and electromagnetic environment that provide effective interaction pathways between processor fields, whereas feedback modifies subsequent local control parameters according to measured or estimated processor state. These mechanisms are architecturally distinct. The present paper does not report an independent coupling-characterization dataset or an open-loop versus closed-loop feedback ablation; processor-level circuit and state measurements are reported for the integrated system.

\FloatBarrier

\subsection{Engineering scale-up path to 9,984 processor-level logical qubits}\label{sec:3.8}

The architecture is designed to scale from the experimentally demonstrated 256-qubit Zeus-256 processor to a processor configuration containing up to 9,984 processor-level logical qubits. Within the architecture presented here, this transition is not presently expected to require a different qubit modality, a new coupling principle, or a new computational-control concept. The core physical and architectural mechanisms required for the scale-up---ensemble-based electron-spin computation, nuclear-spin memory and ancillary control, optical initialization and fluorescence readout, MW/RF pulse control, the programmable biomolecular quantum bus, and processor-level logical-qubit control---are already instantiated in Zeus-256.

Accordingly, the remaining transition from 256 to 9,984 processor-level logical qubits is treated as an engineering scale-up program. The principal tasks are replication and densification of processor fields; expansion of MW/RF and optical control channels; RFSoC I/O and control-bandwidth scaling; packaging and interconnect engineering; calibration automation; optical collection uniformity; channel isolation and crosstalk management; power and thermal management; manufacturing yield; and system-level validation at increased register size. These are substantial engineering tasks and, on the basis of the Zeus-256 architecture, are not presently expected to require a different qubit modality or coupling principle.

The 9,984-qubit figure is therefore presented as an architectural engineering target rather than as an experimentally demonstrated register size. Reaching that target still requires verification that fabrication yield, calibration throughput, field uniformity, coupling selectivity, control bandwidth, and aggregate logical-state performance remain within specification at scale. This distinction is maintained throughout the Results and Discussion sections: Zeus-256 provides the experimental evidence, while Zeus-9984 represents the engineering scale-up of the same processor architecture.

\begin{table}[htbp]
\centering
\small
\setlength{\tabcolsep}{3pt}
\setlength{\tablewidth}{\dimexpr\linewidth-2\tabcolsep\relax}
\renewcommand{\arraystretch}{1.18}
\renewcommand{\thetable}{1}
\caption{Zeus-256 experimental processor architecture.}\label{tab:1}
\begin{tabular}{@{}L{0.310000\tablewidth}L{0.690000\tablewidth}@{}}
\toprule
\textbf{Parameter} & \textbf{Zeus-256 experimental value} \\
\midrule
Diamond material & CVD diamond with NV-centre active population \\
Diamond dimensions & 10 mm \(\times\) 10 mm \(\times\) 0.5 mm \\
Processor field grid & 16 \(\times\) 16 \\
Simultaneously active processor-level logical qubits & 256 \\
Processor-level logical-qubit implementation & NV electron-spin ensemble with associated nuclear-spin memory/ancilla \\
Logical control protocol & Proprietary electron--nuclear-spin control protocol with error-management functions \\
Inter-field coupling / quantum bus & Optically programmable biomolecular quantum-bus coupling fabric with internal metastable-triplet spin mediator \\
Circuit input & OpenQASM 2.0, transpiled to processor-specific pulse and bus-configuration schedules \\
Hardware control & RFSoC; per-field MW/RF frequency, amplitude, phase, and timing \\
Electron-spin control & Individual microwave excitation around calibrated NV resonance \\
Nuclear-spin control & Individual RF excitation of associated nuclear-spin degrees of freedom \\
Initialization/readout & Optical initialization and spin-dependent fluorescence \\
Operating regime & Room temperature \\
\bottomrule
\end{tabular}
\end{table}

\FloatBarrier

\section{Physical Model of the Biomolecular Quantum Bus and Programmable Coupling}\label{sec:4}

\FloatBarrier

\subsection{Electromagnetic mechanism and formation of the effective biomolecular quantum bus}\label{sec:4.1}

The biomolecular quantum-bus material is treated at the processor level as an optically reconfigurable biomolecular spin medium rather than solely as a passive dielectric. The proprietary material uses a protein--chromophore system similar to EYFP, with a metastable electronic triplet spin degree of freedom that can be optically addressed and used internally as a molecular mediator. EYFP itself has independently been demonstrated as an optically addressable fluorescent-protein triplet spin qubit \cite{ref25}; this external result provides a physical precedent for the relevant class of molecular spin degree of freedom but does not identify the proprietary Zeus material. In Zeus-256, optical programming changes the local dielectric response and coupling state of the biomolecular material, altering the boundary conditions and spatial distribution of the microwave and radio-frequency fields used to control the NV electron- and nuclear-spin ensembles. In the programmed state, the biomolecular spin subsystem forms part of an effective bus-mediated interaction channel across selected processor regions. This channel allows quantum information associated with one selected processor field to influence the coherent evolution of another selected field even when the fields are spatially separated. The bus is therefore distinct from both direct bare nearest-neighbour spin exchange and classical feedback.

The relevant constitutive relation is

\[
\mathbf D(\mathbf r,t)=\varepsilon(\mathbf r,t)\,\mathbf E(\mathbf r,t)
\]

where \(\varepsilon(\mathbf r,t)\) denotes the spatially and temporally programmed dielectric response of the coupling layer. Under optical programming,

\[
\varepsilon(\mathbf r,t)=\varepsilon_0(\mathbf r)+\Delta\varepsilon_{\mathrm{opt}}(\mathbf r,t)
\]

Through Maxwell's equations, a non-zero \(\Delta\varepsilon_{\mathrm{opt}}\) changes the local electric and magnetic field solutions generated by the MW/RF control structures. The processor-level consequence is a change in the field amplitude, phase, spatial overlap, and effective control response experienced by selected logical-qubit fields.

For an electron-spin ensemble in processor field i, the local microwave-driven Rabi rate may be expressed generically as

\[
\Omega_i\propto\gamma_e\lvert\mathbf B_1(\mathbf r_i)\rvert
\]

where \(\gamma_e\) is the electron-spin gyromagnetic ratio and \(\mathbf B_1(\mathbf r_i)\) is the local transverse microwave magnetic field. The optically programmed dielectric state changes \(\mathbf B_1(\mathbf r)\) and the corresponding RF field distribution for nuclear-spin control. The experimentally relevant quantity is therefore the transfer from optical programming pattern to local MW/RF control response rather than the undisclosed molecular composition itself \cite{ref12,ref13}.

At the processor-functional level, the biomolecular quantum bus can be understood as a programmable mediator shared by multiple processor fields. The biomolecular spin degree of freedom is an internal component of this mediator and is not exposed as an additional user qubit register. Activating a selected bus configuration establishes an effective interaction pathway between designated fields i and j; the processor control stack then applies synchronized MW/RF and optical control so that quantum information can be transferred or an entangling interaction can be accumulated through that shared channel. Because the interaction is bus-mediated, the participating fields need not be adjacent in the 16 \(\times\) 16 array. The exact protein--chromophore identity, microscopic triplet-state dynamics, bus-transfer Hamiltonian, pulse sequence, and calibration coefficients are proprietary and are intentionally not specified. Accordingly, the manuscript characterizes the bus through the externally observable behavior of the integrated processor---execution of compiled long-range multi-qubit interactions and the generation of Bell and multipartite GHZ correlations---rather than through disclosure of the microscopic bus implementation.

\begin{figure}[htbp]
\centering
\renewcommand{\thefigure}{3}\renewcommand{\theHfigure}{3}
\includegraphics[width=\linewidth,height=0.69\textheight,keepaspectratio,trim={0.0000bp 99.0000bp 0.0000bp 0.0000bp},clip]{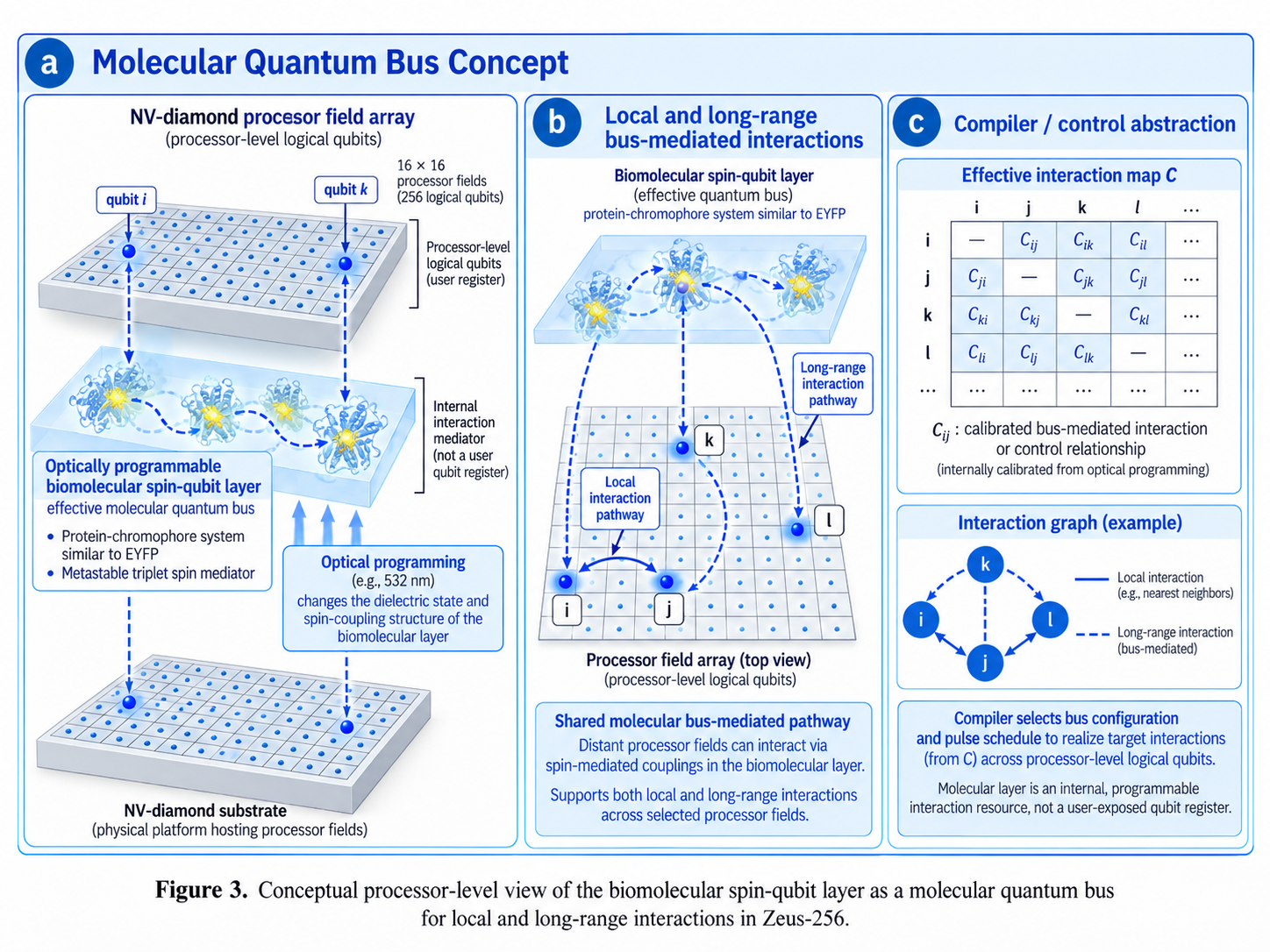}
\caption{Processor-level schematic of the Zeus biomolecular quantum bus. The optically programmable biomolecular layer contains a proprietary protein--chromophore spin system similar to EYFP, incorporating a metastable electronic triplet spin degree of freedom used internally as part of the quantum bus. This molecular spin subsystem is an interaction mediator and is not exposed as an additional processor-level logical-qubit register. Optical programming changes the dielectric and coupling state of the biomolecular layer, while synchronized MW/RF control and fluorescence readout operate on selected NV-diamond processor-level logical qubits. The programmed material supports both local and long-range bus-mediated interaction pathways, allowing spatially separated processor fields to exchange quantum information and participate in mediated entangling operations without requiring a strictly nearest-neighbour-only connectivity model. The compiler represents these available pathways through the proprietary effective interaction map C and selects the corresponding bus configuration and pulse schedule for each compiled circuit. Reference \cite{ref25} is cited only as an external physical precedent for this class of optically addressable triplet-state biomolecular spin degree of freedom; the exact Zeus molecular identity, microscopic Hamiltonian, bus-transfer dynamics, and fabrication layout are not disclosed.}\label{fig:3}
\end{figure}

Figure~\ref{fig:3} summarizes the processor-level interpretation of the biomolecular coupling material and its role as a programmable biomolecular quantum bus. The schematic distinguishes the NV-diamond processor-level logical qubits from the internal protein--chromophore mediator, illustrates local and long-range bus-mediated interactions, and shows the effective interaction-map abstraction used by the compiler. The biomolecular layer is represented as a proprietary protein--chromophore system similar to EYFP, whose metastable electronic triplet spin degree of freedom participates internally in the coupling fabric; the exact molecular identity and microscopic dynamics remain proprietary.

\FloatBarrier

\subsection{Effective representation of local and long-range bus-mediated interactions}\label{sec:4.2}

At the compiler and control level, the processor represents the programmed connectivity through a proprietary calibrated effective interaction map C. An element \(C_{ij}\) denotes the internally calibrated bus-mediated interaction or control relationship available between processor-level logical-qubit fields i and j under a given optical programming configuration. \(C_{ij}\) may therefore describe either a local interaction or a long-range interaction between spatially separated fields. This representation allows the compiler to reason about the biomolecular material as a programmable quantum bus without exposing the microscopic molecular dynamics. C is an internal processor abstraction and does not imply a particular microscopic two-spin Hamiltonian unless such a Hamiltonian is independently established. The detailed calibration data underlying C are not disclosed.

The control stack maps circuit interaction requirements onto proprietary biomolecular-bus programming patterns and field-specific MW/RF pulse schedules. When a circuit requires an interaction between spatially separated processor fields, the compiler selects a bus configuration that exposes the corresponding effective interaction pathway and coordinates the associated control sequence. The bus therefore permits non-nearest-neighbour interactions without requiring every such circuit edge to be decomposed into a chain of nearest-neighbour SWAP operations. Reconfiguration is implemented internally by changing the optical state of the biomolecular material and selecting or recalibrating the corresponding effective interaction map. The detailed mapping procedure, bus-transfer characteristics, intended-versus-unintended field response, selectivity, repeatability, drift, crosstalk, reconfiguration time, and native entangling pulse implementation form part of the proprietary coupling implementation and are not disclosed individually in this work. The present paper evaluates the integrated subsystem at the processor boundary through compiled circuit execution and resulting quantum-state measurements.

\FloatBarrier

\subsection{Distinction from measurement-conditioned feedback}\label{sec:4.3}

The biomolecular quantum bus and measurement-conditioned feedback act at different levels of the processor. The bus combines an internal metastable-triplet molecular spin mediator with optically programmable dielectric/electromagnetic reshaping to configure effective interactions before and during coherent control, whereas feedback uses fluorescence-derived state estimates to modify subsequent control operations. A circuit may use either mechanism or both, and their roles are architecturally distinct. The broader use of measurement backaction, feedback, and engineered non-unitary dynamics as quantum-control resources is established in prior work \cite{ref4,ref8}.

This distinction is important for interpretation of the integrated control stack. In principle, coupling and feedback can be isolated through dedicated ablation experiments; however, independent coupling-characterization and open-loop versus closed-loop feedback datasets are not reported in the present work. Section~\ref{sec:5} therefore evaluates the integrated processor primarily at the observable circuit and state-measurement boundary.

\FloatBarrier

\subsection{Relation to scale-up}\label{sec:4.4}

The same physical description applies to the 256-qubit experimental processor and to the proposed 9,984-qubit engineering scale-up. In the demonstrated Zeus-256 architecture, the biomolecular quantum bus provides the common mechanism for both local and long-range effective interaction pathways; increasing the number of processor fields therefore does not intrinsically require converting the architecture to a nearest-neighbour-only coupling model. Scaling instead requires denser or replicated control structures, additional MW/RF and optical channels, greater bus-configuration and calibration throughput, and tighter management of field uniformity and crosstalk. On the basis of the Zeus-256 architecture, this transition is not presently expected to require a different qubit modality or coupling principle. Validation at increased scale remains necessary to determine whether bus selectivity, stability, control bandwidth, and calibration throughput remain usable as the processor is enlarged.

\FloatBarrier

\section{Experimental Methods}\label{sec:5}

\FloatBarrier

\subsection{Experimental configuration and calibration}\label{sec:5.1}

All experiments reported in this work are performed on the Zeus-256 processor described in Section~\ref{sec:3}. Each experimental session begins with calibration of the 256 active processor fields. Calibration includes optically detected magnetic resonance (ODMR) of the active NV ensembles, determination of the microwave resonance frequencies used for electron-spin control, radio-frequency calibration of the associated nuclear-spin degrees of freedom, optical initialization and fluorescence-readout calibration, and verification of the static bias magnetic field.

Each processor field is calibrated independently because the OpenQASM-to-pulse compiler ultimately addresses the 16 \(\times\) 16 array through field-specific microwave and radio-frequency parameters. The calibration state therefore supplies the frequencies, amplitudes, phases, pulse durations, and timing parameters used by the RFSoC control layer during circuit execution.

\FloatBarrier

\subsection{Logical-qubit construction and control protocol}\label{sec:5.2}

One Zeus-256 processor field implements one processor-level logical qubit. The computational degree of freedom is provided by an ensemble of NV electron spins, while associated nuclear-spin degrees of freedom provide memory and ancillary resources. All 256 processor-level logical qubits are simultaneously active during full-register experiments.

A proprietary electron--nuclear-spin control protocol, including error-management functions, is applied at the logical-qubit control layer. The internal control sequence and implementation details are not disclosed in this work. Accordingly, experimental characterization is reported at the observable processor boundary: logical-state preparation, controlled evolution, readout, stability, and circuit-level state quality. The terms microscopic spin, processor field, and logical qubit are kept distinct throughout the analysis.

\FloatBarrier

\subsection{OpenQASM compilation and RFSoC pulse execution}\label{sec:5.3}

Benchmark circuits are supplied in OpenQASM 2.0 form. Before execution, each circuit is transpiled into a Zeus-specific control program consisting of microwave, radio-frequency, optical, timing, measurement, and biomolecular-bus configuration operations. The RFSoC control layer converts this program into synchronized hardware pulse and control sequences and applies the field-specific control parameters obtained during calibration.

The gate counts reported in the benchmark tables refer to non-measurement gates in the original QASM circuits; final measurement statements are excluded unless explicitly stated otherwise. They do not represent the number of native hardware pulses after transpilation. Consequently, the reported QASM gate count is used as an algorithmic circuit descriptor rather than as a direct measure of the number of physical control operations performed by Zeus-256. Table~\ref{tab:6b} is an exception to the specified original-QASM counting convention: its gate counts are reproduced from the Shor run summary, which does not specify the gate set, decomposition level, or inclusion of measurement operations.

\FloatBarrier

\subsection{Proprietary biomolecular quantum-bus and programmable-coupling implementation}\label{sec:5.4}

Zeus-256 incorporates a proprietary optically programmable biomolecular coupling fabric used by the control stack as an effective quantum bus. The material employs a protein--chromophore architecture similar to EYFP and contains a metastable electronic triplet spin degree of freedom that serves internally as a molecular spin mediator. The EYFP fluorescent-protein spin-qubit result \cite{ref25} provides an external precedent for this class of optically addressable triplet spin system; the exact Zeus protein--chromophore identity is not disclosed. In the processor, the biomolecular spin subsystem is not exposed as an additional computational qubit register. Instead, optical programming establishes the biomolecular-bus configuration, and synchronized MW/RF/optical control realizes the compiled interaction on selected processor fields. The bus is a shared, reconfigurable interaction resource through which local and spatially separated fields can exchange quantum information and participate in mediated entangling operations. The exact molecular composition, microscopic triplet-state dynamics, detailed optical programming protocol, calibrated interaction maps, coupling-transfer characteristics, and native inter-field entangling pulse implementation constitute proprietary implementation details and are not disclosed in this work.

For a compiled two-qubit interaction, the control stack identifies the required processor-field pair, selects the corresponding calibrated bus-mediated interaction entry, programs the biomolecular coupling state, and executes the synchronized control sequence. The same procedure is used for non-nearest-neighbour field pairs; spatial separation within the processor array does not by itself require a chain of intermediate qubits. For multipartite circuits, multiple bus-mediated interactions are scheduled according to the transpiled circuit and hardware constraints. The microscopic exchange process and detailed pulse decomposition remain proprietary, but the processor-level role of the bus is directly testable through the resulting two-qubit correlations, multipartite entanglement, and compiled circuit outputs.

Circuit interaction requirements are compiled into processor-specific biomolecular-bus configurations together with MW/RF/optical control schedules, and integrated functionality is assessed through execution of multi-qubit circuits and the corresponding state- and algorithm-level measurements. The proprietary quantum-bus implementation is distinct from measurement-conditioned feedback: the former mediates the quantum interaction pathway between selected processor fields, whereas the latter modifies subsequent local control actions according to measured or estimated state variables. The experiments reported here therefore characterize the externally observable behavior of the integrated bus-enabled processor rather than providing an independently reproducible description of the proprietary biomolecular coupling mechanism itself.

\FloatBarrier

\subsection{Optical readout and state estimation}\label{sec:5.5}

Logical-qubit measurement is based on spin-dependent fluorescence from the NV ensembles. Optical excitation and fluorescence collection are mapped to the processor-field array, and the acquired signals are conditioned and digitized before being converted into processor-level state estimates.

Where a separate readout timing is supplied, the benchmark tables report fluorescence-data interpretation time rather than the complete physical measurement interval. In this paper, \(T_{\mathrm{interpretation}}\) denotes the time required to interpret already acquired fluorescence data and excludes the physical photon-acquisition interval. It therefore does not by itself constitute complete measurement latency or end-to-end wall-clock latency.

\FloatBarrier

\subsection{Multipartite-entanglement validation}\label{sec:5.6}

The entanglement benchmark family prepares Greenberger--Horne--Zeilinger (GHZ) states \cite{ref14} on register sizes extending to the full 256-qubit Zeus-256 array. The specific 256-qubit validation circuit is expressed in OpenQASM 2.0 using a 256-qubit register q[256] and a 256-bit classical register c[256]. The circuit applies H to q[0], followed by CX(q[0], q[i]) for every \(i=1,\ldots,255\), and then measures all qubits. At the QASM level this consists of one Hadamard gate and 255 controlled-NOT gates, for a total of 256 gates. These fan-out interactions connect q[0] to processor fields distributed across the physical array; the compiler therefore maps both local and spatially separated q[0]--q[i] interactions onto the proprietary biomolecular quantum-bus layer. The GHZ experiment consequently exercises the bus as a full-register entanglement resource rather than as a nearest-neighbour-only coupling mechanism. The ideal pre-measurement target is the 256-qubit GHZ state shown in the displayed equation below. A schematic representation of the full-register circuit is provided in Figure~\ref{fig:7}.

\[
|\mathrm{GHZ}_{256}\rangle=\frac{|0\rangle^{\otimes256}+|1\rangle^{\otimes256}}{\sqrt{2}}
\]

Entanglement-state validation is performed separately from the 10-shot timing benchmark. For the full-register 256-qubit experiment, GHZ-state fidelity was evaluated with a population-and-parity protocol rather than full density-matrix tomography \cite{ref20,ref21}. The target-state fidelity was evaluated as:

\[
F_{\mathrm{GHZ}}=\langle\mathrm{GHZ}_{n}|\rho|\mathrm{GHZ}_{n}\rangle=\frac{P+C}{2}
\]

Population P was estimated from 15,000 computational-basis measurements as:

\[
P=p(0^n)+p(1^n)
\]

The target-phase coherence C was obtained from a further 15,000 shots distributed across K = 257 phase settings \(\varphi_k=k\pi/257\), \(k=0,\ldots,256\). Each setting k = 0,\(\ldots\),93 used 59 shots, and each setting k = 94,\(\ldots\),256 used 58 shots. At each phase \(\varphi\), the measured parity was calculated as:

\[
\Pi(\varphi)=\frac{N_{\mathrm{even}}(\varphi)-N_{\mathrm{odd}}(\varphi)}{N_{\mathrm{shots}}(\varphi)}
\]

where \(N_{\mathrm{even}}\) and \(N_{\mathrm{odd}}\) denote outcomes containing an even or odd number of measured ones. The 256th harmonic of the measured parity was represented as:

\[
\Pi_{256}(\varphi)=A\cos(256\varphi)+B\sin(256\varphi)=R\cos(256\varphi+\varphi_0).
\]

For Zeus-256, n = 256. Equal-setting Fourier projection gives \[ A=(2/K)\sum_k\Pi_k\cos(256\varphi_k)\quad\text{and}\quad B=(2/K)\sum_k\Pi_k\sin(256\varphi_k),\] with \(R=\sqrt{A^2+B^2}\) and \(\varphi_0=\operatorname{atan2}(-B,A)\). For the defined positive-phase GHZ target, the coherence entering \(F_{\mathrm{GHZ}}=(P+C)/2\) is C = A = \(R\cos\varphi_0\). The stored preparation phase is 0 rad; the primary fidelity uses the signed cosine component without post-hoc phase optimization. The distinct best-phase estimate is \((P+R)/2\). All reported components and uncertainties are reproduced from the supplied counts in Section~\ref{sec:6.3.1}. This phase-sensitive procedure avoids full density-matrix tomography \cite{ref20,ref21}.

\FloatBarrier

\subsection{Circuit benchmark suite}\label{sec:5.7}

The completed experimental benchmark suite spans multiple circuit classes in order to avoid characterizing processor behavior from a single circuit family. The tested workloads include quantum Fourier transform (QFT) circuits, GHZ entanglement circuits, Grover search \cite{ref15}, graph-state circuits, variational quantum eigensolver (VQE) circuits, quantum neural network (QNN) circuits, arithmetic and miscellaneous QASM benchmarks, portfolio-VQE circuits, travelling-salesman circuits, and routing circuits. The completed suite also includes the Shor-family factorization staircase in Section~\ref{sec:6.6}, based on a Gidney--Ekerå-inspired Ekerå--Håstad implementation with windowed exact arithmetic \cite{ref19,ref24,ref26}. It covers 11 integer sizes from 8 to 84 bits and records 10 attempts per instance. The Grover benchmark used a three-qubit OpenQASM 2.0 circuit targeting output bitstring 101 (reported in c[2]c[1]c[0] order). After preparing the uniform superposition with H on all three qubits, the circuit applies two complete oracle--diffusion iterations. At the QASM level, excluding final measurement operations, the circuit contains 43 gates: 23 Hadamard gates, 16 X gates, and four Toffoli (CCX) gates. Three single-qubit measurements then map q[0], q[1], and q[2] to c[0], c[1], and c[2], respectively.

The benchmark set includes circuits with up to 256 processor-level logical qubits and includes both comparatively shallow full-register circuits and substantially deeper arithmetic circuits. For each original-QASM benchmark entry, the reported qubit count denotes the number of processor-level Zeus-256 qubits addressed by the QASM circuit, while the reported gate count denotes the number of non-measurement gates in the original QASM representation; final measurement statements are excluded unless explicitly stated otherwise.

The full-register qft\_n256 benchmark uses the supplied OpenQASM 2.0 implementation of the exact 256-qubit coherent quantum Fourier transform (QFT) \cite{ref16,ref22,ref23}. Griffiths and Niu \cite{ref16} provide a semiclassical QFT formulation; the circuit executed here is the conventional coherent gate-network form \cite{ref22}. For the full-register qft\_n256 experiment, the input state was randomly assigned before execution rather than fixed to the default all-zero state. For each target qubit q[k], the circuit applies H(q[k]) and controlled-U1 phase rotations from every higher-index qubit q[j] with j > k, using phase angle \(\theta_{j,k}=\pi/2^{j-k}\). After the triangular controlled-phase network, 128 SWAP gates reverse the qubit order through SWAP(q[k],q[255\(-\)k]) for \(k=0,\ldots,127\), followed by a register-wide measurement. Excluding the final measurement statement, the QASM circuit contains 256 Hadamard gates, 32,640 controlled-U1 gates, and 128 SWAP gates, for a total of 33,024 algorithmic gates. The dense pairwise controlled-phase structure makes qft\_n256 a useful full-register compiled-interaction and scheduling stress test. The gate composition is summarized below.

\[
\begin{gathered}N_H=256,\quad N_{\mathrm{CU1}}=\frac{256\cdot255}{2}=32{,}640,\quad N_{\mathrm{SWAP}}=128,\\N_{\mathrm{gates}}=256+32{,}640+128=33{,}024.\end{gathered}
\]

\FloatBarrier

\subsection{Shot count and timing definitions}\label{sec:5.8}

Unless otherwise stated, every entry in the completed original-QASM timing benchmark tables (Tables \ref{tab:3}--\ref{tab:5} and \ref{tab:7}--\ref{tab:12}) is executed using 10 shots. For these tables, the reported hardware execution wall time, \(T_{\mathrm{hw}}\), is the recorded post-transpilation processor-control execution time for the 10-shot benchmark sequence. It excludes circuit transpilation, physical fluorescence photon acquisition, subsequent fluorescence-data interpretation, queueing, and host-side communication; those contributions are not included in \(T_{\mathrm{hw}}\). The completed Shor staircase in Table~\ref{tab:6b} instead records 10 factorization attempts per instance and one execution-time value per row. The supplied summary does not identify that value as a total, mean, or individual-attempt duration, or define its timing boundaries. It is therefore retained as reported, without conversion to a per-shot or per-attempt latency; separate fluorescence-interpretation and classical post-processing times are not supplied.

The reported fluorescence-data interpretation quantity, \(T_{\mathrm{interpretation}}\), is the time required to interpret the already acquired fluorescence data. It excludes the physical photon-acquisition interval. The timing quantities reported in the benchmark tables should therefore be interpreted as

\[
T_{\mathrm{hw}}\equiv\text{post-transpilation processor wall time for 10 shots}
\]

and

\[
T_{\mathrm{interpretation}}=\text{fluorescence data interpretation time}
\]

Neither quantity should be interpreted independently as complete end-to-end application latency. Where total wall-clock latency is required, transpilation, photon acquisition, host communication, queueing, and other system-level contributions must be reported separately. Numerical timing values are reproduced at the precision returned by the control/acquisition stack; the final displayed decimal place should not be interpreted as a formal uncertainty or independently established timer resolution. Because run-to-run timing variance is not reported in the present dataset, comparisons across circuit widths are treated as descriptive timing observations rather than fitted scaling laws.

\FloatBarrier

\subsection{Experimental controls and reporting}\label{sec:5.9}

Hardware calibration measurements, circuit-execution measurements, entanglement-validation measurements, and processor-boundary evaluation of the integrated coupling/control subsystem are treated as distinct reporting classes. The architecture also incorporates measurement-conditioned control, but the present work does not report a dedicated open-loop versus closed-loop feedback ablation. This separation avoids attributing an observed processor-level effect to programmable coupling, feedback, the logical control/error-management protocol, or local pulse control without supporting data.

For circuit benchmarks, the QASM circuit, processor-level qubit count, original QASM gate count, shot count, hardware execution wall time, and fluorescence-data interpretation time are reported explicitly. State-quality measurements are reported separately from timing measurements. Timing data alone are not used as evidence of quantum-state fidelity or quantum computational advantage.

\FloatBarrier

\section{Results}\label{sec:6}

\FloatBarrier

\subsection{Zeus-256 processor configuration}\label{sec:6.1}

The experimental Zeus-256 processor operated at room temperature with a 16 \(\times\) 16 array of 256 simultaneously active processor-level logical qubits. Each processor field implements an electron--nuclear spin processor-level logical-qubit system in which NV electron-spin ensembles provide the driven computational interface and associated nuclear-spin degrees of freedom provide memory and ancillary resources. All 256 processor-level logical qubits are physically active concurrently; the register does not rely on time multiplexing or qubit reuse.

Processor-level characterization in this work is therefore based on demonstrated logical-register operation, quantum-state validation, circuit execution, and algorithm-level measurements, as reported in the following sections.

\begin{table}[htbp]
\centering
\small
\setlength{\tabcolsep}{3pt}
\setlength{\tablewidth}{\dimexpr\linewidth-2\tabcolsep\relax}
\renewcommand{\arraystretch}{1.18}
\renewcommand{\thetable}{2}
\caption{Zeus-256 processor configuration.}\label{tab:2}
\begin{tabular}{@{}L{0.310000\tablewidth}L{0.690000\tablewidth}@{}}
\toprule
\textbf{Metric} & \textbf{Zeus-256} \\
\midrule
Operating regime & Room temperature \\
Processor-field array & 16 \(\times\) 16 \\
Simultaneously active processor-level logical qubits & 256 \\
Processor-level logical-qubit implementation & NV electron-spin ensemble + nuclear-spin memory/ancilla \\
Circuit interface & OpenQASM 2.0 \\
Hardware execution & RFSoC-controlled MW/RF/optical pulse and bus-configuration sequences \\
Connectivity / quantum bus & Optically programmable biomolecular quantum bus with internal triplet-spin mediator \\
\bottomrule
\end{tabular}
\end{table}

\FloatBarrier

\subsection{Bell--CHSH two-qubit entanglement verification}\label{sec:6.2}

A Bell-pair experiment was used as an independent two-qubit verification of nonclassical correlations on the Zeus processor. Measurements used 1,024 shots per CHSH setting. In the Z/Z basis, 542 of 1,024 observations yielded 00 and 482 yielded 11, while the 01 and 10 outcomes were absent in this acquisition. Using the CHSH formulation \cite{ref17}, the four correlation measurements produced S = 2.8008 \(\pm\) 0.0446 (one-sample standard error), compared with the local-classical bound S = 2 and the Tsirelson quantum bound \(2\sqrt{2}\) \(\approx\) 2.8284 \cite{ref18}.

\[
\begin{gathered}S=E(A_0,B_0)+E(A_0,B_1)+E(A_1,B_0)-E(A_1,B_1),\\ |S|\leq 2\quad\text{(local classical)},\qquad |S|\leq 2\sqrt{2}\quad\text{(quantum)}.\end{gathered}
\]

The measured value exceeds the classical bound by approximately 18 standard errors under the displayed uncertainty estimate. A 20,000-resample bootstrap analysis produced mean S = 2.80131 with \(\sigma\) = 0.04442 and a distribution centred close to the directly measured value. These measurements provide processor-level evidence of Bell-state entanglement. The experiment is not presented here as a loophole-free Bell test.

\begin{figure}[htbp]
\centering
\renewcommand{\thefigure}{4}\renewcommand{\theHfigure}{4}
\includegraphics[width=\linewidth,height=0.69\textheight,keepaspectratio]{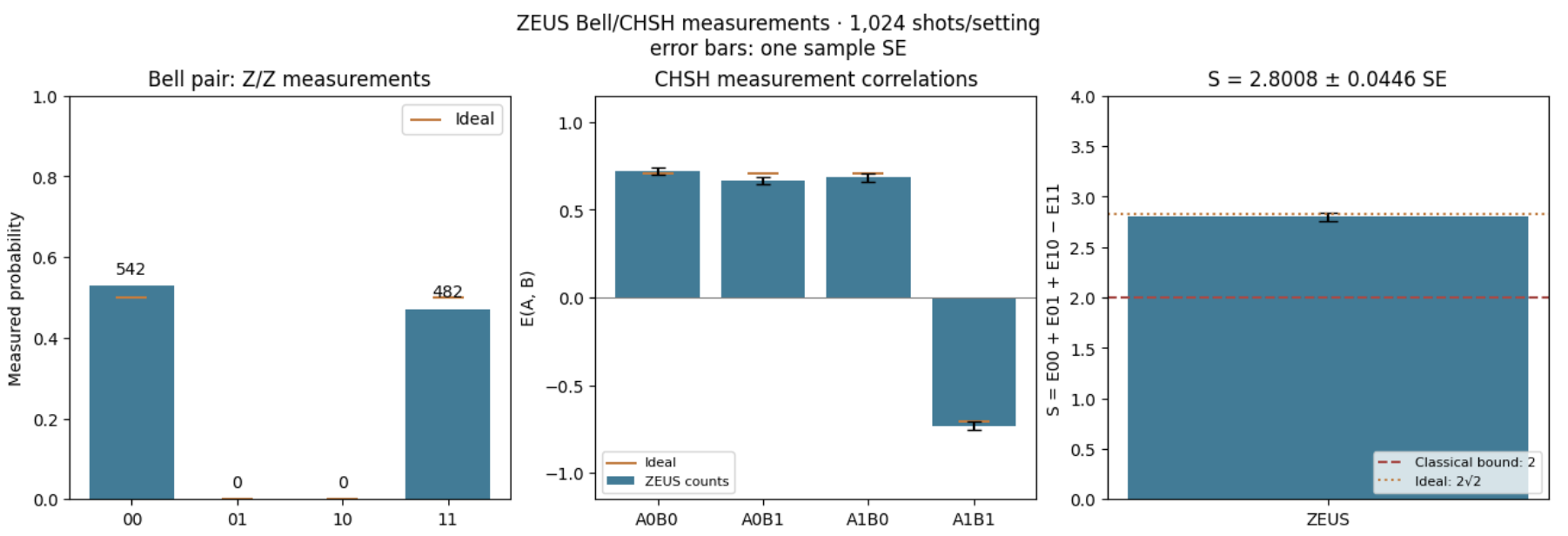}
\caption{Bell--CHSH measurement results for Zeus. Left: Z/Z Bell-pair output probabilities. Centre: measured CHSH correlators. Right: measured CHSH parameter relative to the classical and Tsirelson bounds. Error bars denote one-sample standard errors.}\label{fig:4}
\end{figure}

\begin{figure}[htbp]
\centering
\renewcommand{\thefigure}{5}\renewcommand{\theHfigure}{5}
\includegraphics[width=\linewidth,height=0.69\textheight,keepaspectratio]{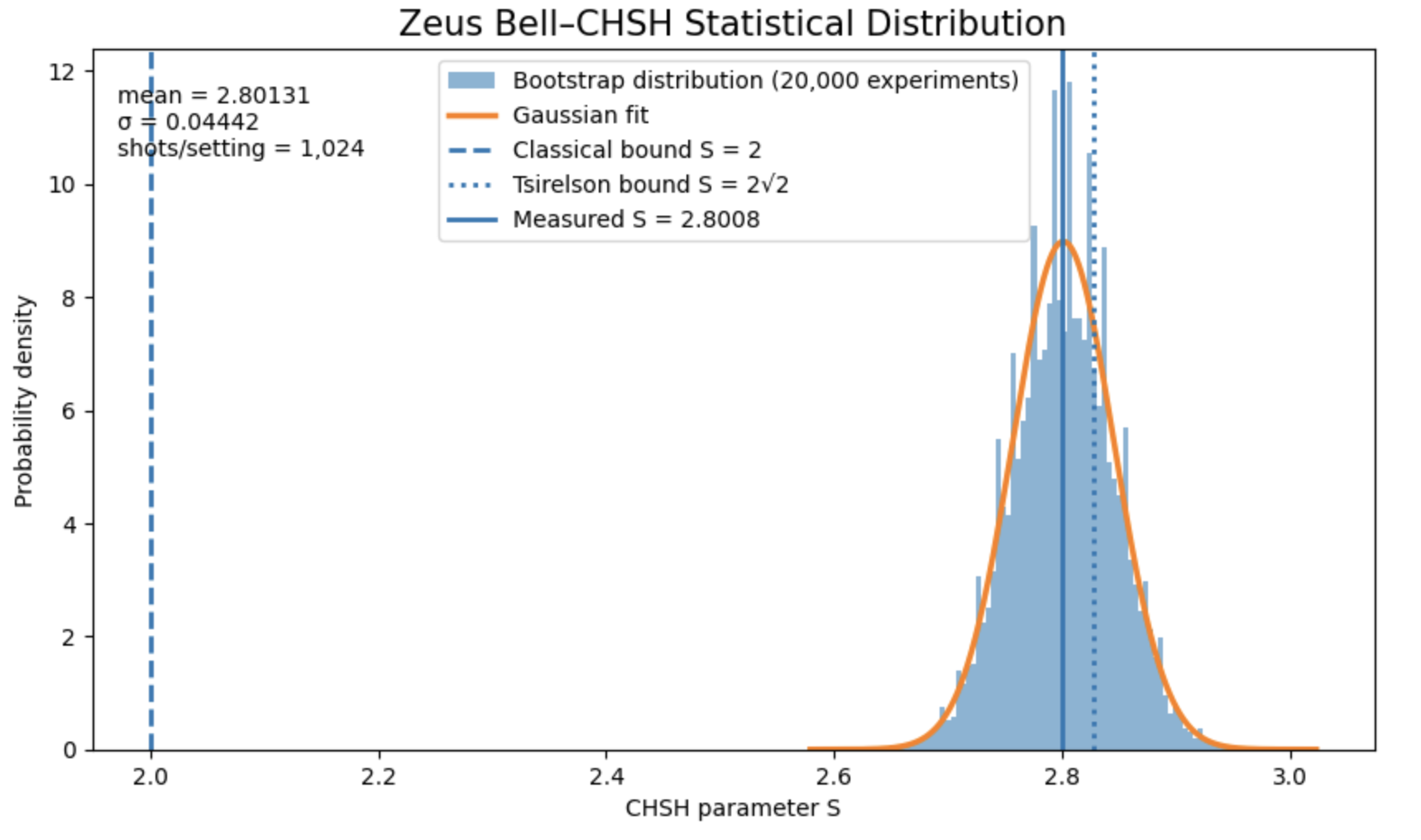}
\caption{Bootstrap distribution of the Zeus CHSH parameter using 20,000 resampled experiments. The distribution has mean 2.80131 and \(\sigma\) = 0.04442; the directly measured value is S = 2.8008.}\label{fig:5}
\end{figure}

\FloatBarrier

\subsection{Multipartite GHZ-state experiments}\label{sec:6.3}

The entanglement benchmark family prepared Greenberger--Horne--Zeilinger (GHZ) states \cite{ref14} across register sizes from 21 to 256 processor-level logical qubits. The 256-qubit circuit used the full simultaneously active Zeus-256 register and consisted, at the QASM level, of H(q[0]) followed by 255 fan-out controlled-NOT operations CX(q[0],q[i]) for \(i=1,\ldots,255\), before measurement of all 256 qubits. Because q[0] is coupled algorithmically to fields distributed throughout the 16 \(\times\) 16 array, the full-register GHZ circuit requires interactions that extend beyond nearest-neighbour field pairs. These interactions are mapped by the compiler onto the proprietary biomolecular quantum bus, providing processor-boundary evidence that the bus-enabled architecture supports long-range interaction pathways as part of multipartite entanglement generation. The ideal pre-measurement target is the 256-qubit GHZ state shown in the displayed equation below. The corresponding schematic circuit structure is shown in Figure~\ref{fig:7}.

\[
|\mathrm{GHZ}_{256}\rangle=\frac{|0\rangle^{\otimes256}+|1\rangle^{\otimes256}}{\sqrt{2}}
\]

The timing results in Table~\ref{tab:3} correspond to separate 10-shot executions and are reported independently of the higher-statistics state-validation experiment.

For the full-register state-validation experiment, the 256-qubit GHZ circuit was characterized using a population-and-parity fidelity protocol \cite{ref20,ref21} with 30,000 total measurement shots: 15,000 shots in the computational Z basis and 15,000 shots distributed across 257 phase-rotated basis settings. The Z-basis data yielded the GHZ population:

\[
P=p(0^{256})+p(1^{256})
\]

For the coherence measurement, parity was evaluated at each phase setting as:

\[
\Pi(\varphi)=\frac{N_{\mathrm{even}}(\varphi)-N_{\mathrm{odd}}(\varphi)}{N_{\mathrm{shots}}(\varphi)}
\]

The 256th Fourier harmonic of the phase-dependent parity was reconstructed as:

\[
\Pi_{256}(\varphi)=R\cos(256\varphi+\varphi_0).
\]

where R = 0.997962328 is the coherence magnitude and \(\varphi_0\) = 0.009203968 rad is the phase offset in this convention. Fidelity to the fixed positive-phase target uses C = A = \(R\cos\varphi_0\) = 0.997920058. The experimentally estimated \(\mathrm{GHZ}^{+}\) fidelity was then calculated as:

\[
F_{\mathrm{GHZ}}=\frac{P+C}{2}=\langle\mathrm{GHZ}_{256}|\rho|\mathrm{GHZ}_{256}\rangle
\]

The fixed-phase \(\mathrm{GHZ}^{+}\) fidelity was \(F_{\mathrm{GHZ}}\) = 0.998960029, or 99.8960\%, with an estimated standard error of 0.002885024 (0.2885 percentage points). The conservative 95\% Hoeffding interval is [0.982341653, 1.015578405] before intersection with the physical range, giving [0.982341653, 1]. No full density-matrix reconstruction was performed. This 30,000-shot state-validation experiment is distinct from the 10-shot timing entries in Table~\ref{tab:3}.

\subsubsection{Detailed GHZ fidelity reporting}\label{sec:6.3.1}

The complete acquisition contains 258 settings and 30,000 recorded shots: one Z-basis setting with 15,000 shots and 257 rotated-basis settings with 15,000 shots in total. The Z-basis counts were 7,526 all-zero outcomes and 7,474 all-one outcomes, with no other bitstrings recorded. Thus p(\(0^{256}\)) = 0.501733333, p(\(1^{256}\)) = 0.498266667, and P = 1.000000000. Supplementary Data 1 (\texttt{ghz\_counts\_fidelity.json}) contains the complete 256-bit count dictionaries, exact phases, shot allocations, backend label, and stored result fields. Supplementary Note 1 (\texttt{GHZ-256.md}) contains the acquisition log and numerical summary; these two files are available in the supplementary repository identified under Data and code availability.

\begin{table}[htbp]
\centering
\small
\setlength{\tabcolsep}{3pt}
\setlength{\tablewidth}{\dimexpr\linewidth-8\tabcolsep\relax}
\renewcommand{\arraystretch}{1.18}
\renewcommand{\thetable}{S1}
\captionsetup{name=Supplementary Table}
\caption{GHZ-256 fidelity components. Uncertainties are estimated standard errors from independent-shot counting statistics; those for R and \(\varphi_0\) use first-order propagation from the joint A,B covariance.}\label{tab:S1}
\begin{tabular}{@{}L{0.240000\tablewidth}C{0.100000\tablewidth}L{0.250000\tablewidth}R{0.205000\tablewidth}R{0.205000\tablewidth}@{}}
\toprule
\textbf{Quantity} & \textbf{Symbol} & \textbf{Estimator} & \textbf{Value} & \textbf{Estimated SE} \\
\midrule
Z-basis shots & \(N_Z\) & Recorded shots & 15,000 & --- \\
Rotated shots & \(N_{\mathrm{rot}}\) & Recorded shots & 15,000 & --- \\
Phase settings & K & \(\varphi_k=k\pi/257\) & 257 & --- \\
GHZ population & P & \(p(0^{256})+p(1^{256})\) & 1.000000000 & 0 (plug-in) \\
Coherence magnitude & R & \(\sqrt{A^2+B^2}\) & 0.997962328 & 0.005769085 \\
Phase offset & \(\varphi_0\) & \(\operatorname{atan2}(-B,A)\) & 0.009203968 rad & 0.010077186 rad \\
Target-phase coherence & C = A & \(R\cos\varphi_0\) & 0.997920058 & 0.005770047 \\
\(\mathrm{GHZ}^{+}\) fidelity & \(F_{\mathrm{GHZ}}\) & \((P+A)/2\) & 0.998960029 & 0.002885024 \\
\bottomrule
\end{tabular}
\end{table}

\begin{table}[htbp]
\centering
\small
\setlength{\tabcolsep}{3pt}
\setlength{\tablewidth}{\dimexpr\linewidth-6\tabcolsep\relax}
\renewcommand{\arraystretch}{1.18}
\renewcommand{\thetable}{S2}
\captionsetup{name=Supplementary Table}
\caption{GHZ-256 computational-basis population measurements. The observed zero leakage does not imply zero uncertainty in the underlying population.}\label{tab:S2}
\begin{tabular}{@{}L{0.270000\tablewidth}R{0.130000\tablewidth}R{0.220000\tablewidth}L{0.380000\tablewidth}@{}}
\toprule
\textbf{Outcome class} & \textbf{Count} & \textbf{Probability} & \textbf{Notes} \\
\midrule
\(0^{256}\) & 7,526 & 0.501733333 & All-zero outcome \\
\(1^{256}\) & 7,474 & 0.498266667 & All-one outcome \\
All other strings & 0 & 0.000000000 & No leakage recorded \\
Total & 15,000 & 1.000000000 & Z-basis shots \\
GHZ population & 15,000 & 1.000000000 & Sum of target populations \\
\bottomrule
\end{tabular}
\end{table}

\begin{table}[htbp]
\centering
\small
\setlength{\tabcolsep}{3pt}
\setlength{\tablewidth}{\dimexpr\linewidth-12\tabcolsep\relax}
\renewcommand{\arraystretch}{1.18}
\renewcommand{\thetable}{S3}
\captionsetup{name=Supplementary Table}
\caption{Selected GHZ-256 phase-resolved parity measurements and totals over all 257 settings. Index k is zero-based; \(\varphi_k=k\pi/257\). The complete phase-resolved 256-bit count record is supplied in Supplementary Data 1.}\label{tab:S3}
\begin{tabular}{@{}C{0.120000\tablewidth}R{0.170000\tablewidth}R{0.110000\tablewidth}R{0.110000\tablewidth}R{0.110000\tablewidth}R{0.190000\tablewidth}R{0.190000\tablewidth}@{}}
\toprule
\textbf{Setting k} & \textbf{\(\varphi_k\) (rad)} & \textbf{Shots} & \textbf{Even} & \textbf{Odd} & \textbf{Parity \(\Pi_k\)} & \textbf{SE(\(\Pi_k\))} \\
\midrule
0 & 0.000000 & 59 & 59 & 0 & 1.000000 & 0.000000 \\
1 & 0.012224 & 59 & 0 & 59 & -1.000000 & 0.000000 \\
2 & 0.024448 & 59 & 59 & 0 & 1.000000 & 0.000000 \\
32 & 0.391171 & 59 & 59 & 0 & 1.000000 & 0.000000 \\
64 & 0.782342 & 59 & 51 & 8 & 0.728814 & 0.089907 \\
93 & 1.136841 & 59 & 23 & 36 & -0.220339 & 0.128079 \\
94 & 1.149065 & 58 & 42 & 16 & 0.448276 & 0.118399 \\
128 & 1.564684 & 58 & 27 & 31 & -0.068966 & 0.132138 \\
192 & 2.347026 & 58 & 6 & 52 & -0.793103 & 0.080676 \\
256 & 3.129369 & 58 & 0 & 58 & -1.000000 & 0.000000 \\
All 257 & --- & 15,000 & 7,523 & 7,477 & --- & --- \\
\bottomrule
\end{tabular}
\end{table}

\begin{table}[htbp]
\centering
\small
\setlength{\tabcolsep}{3pt}
\setlength{\tablewidth}{\dimexpr\linewidth-6\tabcolsep\relax}
\renewcommand{\arraystretch}{1.18}
\renewcommand{\thetable}{S4}
\captionsetup{name=Supplementary Table}
\caption{GHZ-256 estimation and uncertainty summary. Fidelity is reported relative to the fixed \(\mathrm{GHZ}^{+}\) target; phase optimization is reported separately.}\label{tab:S4}
\begin{tabular}{@{}L{0.270000\tablewidth}R{0.180000\tablewidth}R{0.180000\tablewidth}L{0.370000\tablewidth}@{}}
\toprule
\textbf{Quantity / method} & \textbf{Value} & \textbf{Estimated SE} & \textbf{95\% interval / notes} \\
\midrule
Coherence magnitude R & 0.997962328 & 0.005769085 & First-order propagation \\
Phase offset \(\varphi_0\) (rad) & 0.009203968 & 0.010077186 & First-order propagation \\
Target coherence A & 0.997920058 & 0.005770047 & Signed cosine component \\
Population P & 1.000000000 & 0 (plug-in) & Exact binomial: [0.999754105, 1] \\
Fixed-phase fidelity & 0.998960029 & 0.002885024 & Hoeffding, physical: [0.982341653, 1] \\
Phase-optimized fidelity & 0.998981164 & --- & \((P+R)/2\); diagnostic only \\
Harmonic description & \(R^2\) = 0.984698 & --- & Unweighted residual statistic; 257 settings \\
\bottomrule
\end{tabular}
\end{table}

\textbf{Estimation and uncertainty.} The discrete projection gives A = 0.9979200577904833 and B = \(-\)0.0091850839562589. Equal weight is assigned to each phase setting, using its measured parity mean; the 58/59-shot allocation enters the variance rather than the Fourier weights. For independent \(\pm\)1 outcomes, \(v_k\) estimates Var(\(\Pi_k\)). The projection vector \(u_k\) and the estimated covariance of (A, B) are defined by:

\[
\begin{gathered}v_k=\frac{1-\Pi_k^2}{N_k-1},\qquad u_k=\frac{2}{K}\bigl[\cos(256\varphi_k),\,\sin(256\varphi_k)\bigr]^{\mathsf T},\\\widehat{\operatorname{Cov}}\!\bigl[(A,B)^{\mathsf T}\bigr]=\sum_k v_k u_k u_k^{\mathsf T}.\end{gathered}
\]

This gives SE(A) = 0.005770047 and SE(B) = 0.010056100. The fidelity standard error is:

\[
\operatorname{SE}(F_{\mathrm{GHZ}})=\frac{1}{2}\sqrt{\operatorname{Var}(P)+\operatorname{Var}(A)}=0.002885024.
\]

Because every Z-basis shot belongs to the target population, the plug-in Var(P) is zero; the two-sided exact 95\% binomial interval for P is [0.999754105, 1].

The reported conservative interval uses Hoeffding's inequality applied directly to the independent bounded shot contributions to \(F_{\mathrm{GHZ}}\). Its half-width is:

\[
\Delta=\sqrt{\frac{\ln(40)}{2}\left[\frac{1}{4N_Z}+\frac{4}{K^2}\sum_k\frac{\cos^2(256\varphi_k)}{N_k}\right]}=0.016618376.
\]

The resulting raw interval [0.982341653, 1.015578405] is retained in Supplementary Data 1; intersection with \(0\leq F_{\mathrm{GHZ}}\leq1\) gives [0.982341653, 1]. This bound includes finite-sample population uncertainty even when the population plug-in variance is zero. The standard errors and confidence interval quantify counting uncertainty under independent trials and stable measurement settings; they do not include uncharacterized calibration drift or readout systematics.

\begin{figure}[htbp]
\centering
\renewcommand{\thefigure}{S1}\renewcommand{\theHfigure}{S1}
\captionsetup{name=Supplementary Figure}
\includegraphics[width=\linewidth,height=0.69\textheight,keepaspectratio]{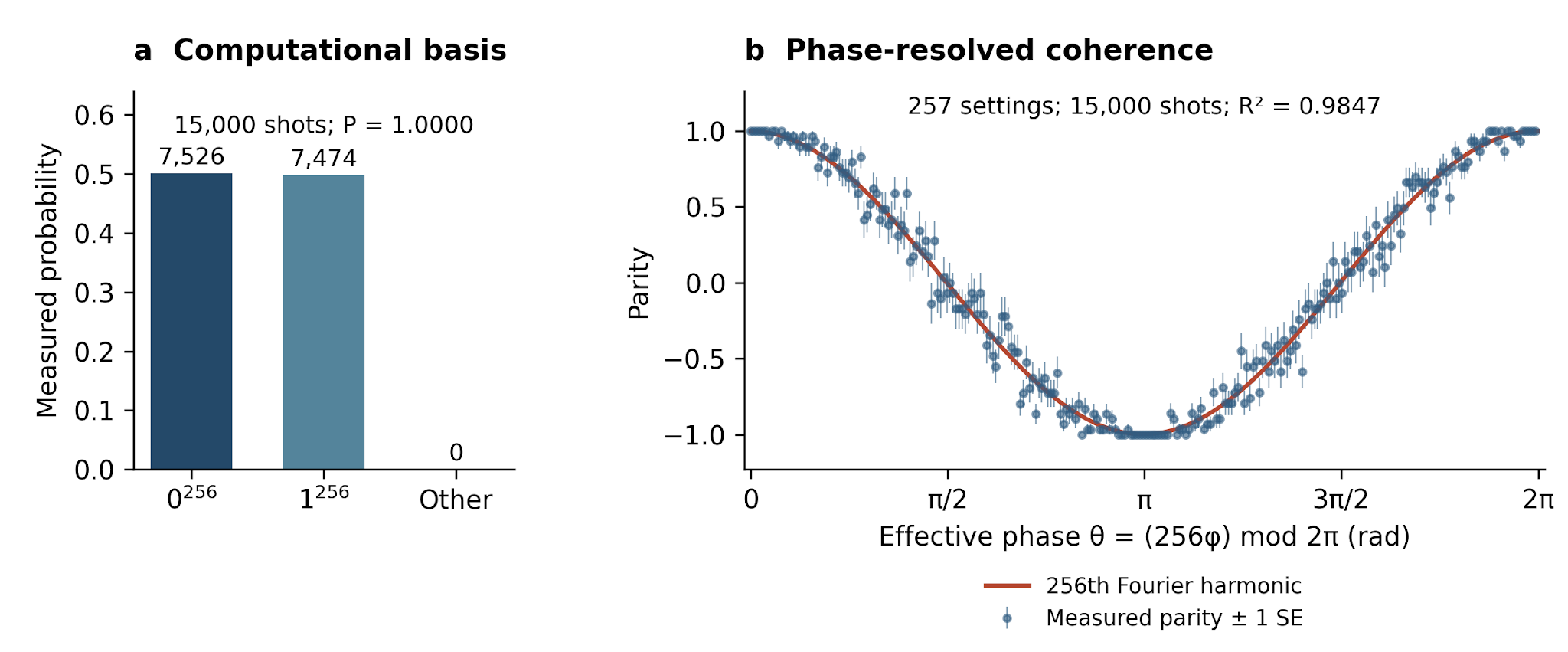}
\caption{GHZ-256 population and parity measurements. (a) Computational-basis outcome probabilities from 15,000 shots. (b) All 257 measured parities, reordered by effective phase \(\theta=(256\varphi)\bmod 2\pi\) without binning or averaging, with estimated \(\pm\)1 SE error bars using \((1-\Pi_k^2)/(N_k-1)\). The line is \(R\cos(\theta+\varphi_0)\), with R = 0.997962328 and \(\varphi_0\) = 0.009203968 rad. The unweighted descriptive statistic is \(R^2\) = 0.984698; it is not used to obtain the fidelity confidence interval.}\label{fig:S1}
\end{figure}

The fixed-target fidelity \(F_{\mathrm{GHZ}}\) = 0.998960029 corresponds to 99.8960\% \(\pm\) 0.2885 percentage points (estimated SE). Optimizing the target phase would give 0.998981164 (99.8981\%); that value is reported only as a diagnostic and is not substituted for the fixed-phase result. Under the trusted local-measurement model and independent-shot assumptions, the 95\% lower fidelity bound of 0.982341653 exceeds the GHZ biseparability threshold of 1/2, supporting genuine multipartite entanglement across the 256 addressed qubits \cite{ref21}. All recorded counts are used, without additional filtering or readout/SPAM correction in this analysis. The supplied files do not document upstream readout mitigation, rejected shots, or a separate phase-compensation calibration; such procedures are therefore not inferred from these records.

\begin{table}[htbp]
\centering
\small
\setlength{\tabcolsep}{3pt}
\setlength{\tablewidth}{\dimexpr\linewidth-8\tabcolsep\relax}
\renewcommand{\arraystretch}{1.18}
\renewcommand{\thetable}{3}
\caption{GHZ entanglement circuit benchmarks on Zeus-256.}\label{tab:3}
\begin{tabular}{@{}L{0.340000\tablewidth}R{0.075000\tablewidth}R{0.105000\tablewidth}R{0.250000\tablewidth}R{0.230000\tablewidth}@{}}
\toprule
\textbf{Problem} & \textbf{Qubits} & \textbf{QASM gates} & \textbf{Hardware execution wall time (ms; 10 shots)} & \textbf{Fluorescence interpretation (ms)} \\
\midrule
entanglement\_n21 & 21 & 21 & 0.8844 & 0.2015 \\
entanglement\_n22 & 22 & 22 & 0.9026 & 0.2043 \\
entanglement\_n23 & 23 & 23 & 0.9466 & 0.2069 \\
entanglement\_n27 & 27 & 27 & 1.0515 & 0.2221 \\
entanglement\_n28 & 28 & 28 & 1.1858 & 0.2213 \\
entanglement\_n29 & 29 & 29 & 1.2153 & 0.2250 \\
entanglement\_n63 & 63 & 63 & 2.2218 & 0.3270 \\
entanglement\_n64 & 64 & 64 & 2.3141 & 0.3312 \\
entanglement\_n127 & 127 & 127 & 4.1300 & 0.5134 \\
entanglement\_n128 & 128 & 128 & 4.1389 & 0.5152 \\
entanglement\_n255 & 255 & 255 & 9.5191 & 0.8943 \\
entanglement\_n256 & 256 & 256 & 9.6644 & 0.8926 \\
\bottomrule
\end{tabular}
\end{table}

\FloatBarrier

\subsection{Quantum Fourier transform timing across circuit widths}\label{sec:6.4}

QFT circuits provide a systematic width series from 12 to 256 processor-level logical qubits \cite{ref16,ref22}. The supplied qft\_n256 circuit is the exact 256-qubit coherent QFT in OpenQASM 2.0 \cite{ref23}: 256 H gates, 32,640 controlled-U1 phase rotations, and 128 final SWAP operations, for 33,024 QASM gates excluding the final register-wide measurement statement. The full-register qft\_n256 experiment used a randomly assigned input state rather than a fixed all-zero input. Its triangular controlled-phase network contains algorithmic interactions between every qubit and all higher-index qubits, followed by qubit-order reversal. The full-register circuit therefore exercises all 256 Zeus processor-level logical qubits and presents a dense nonlocal interaction pattern to the compiler. Across the QFT width series, the original QASM gate count ranges from 84 to 33,024 and the recorded 10-shot hardware execution wall times range from 0.1621 ms to 3.1459 ms. The values are descriptive observations and are not interpreted as a monotonic scaling law. The circuit structure is illustrated in Figure~\ref{fig:9}.

\begin{table}[htbp]
\centering
\small
\setlength{\tabcolsep}{3pt}
\setlength{\tablewidth}{\dimexpr\linewidth-8\tabcolsep\relax}
\renewcommand{\arraystretch}{1.18}
\renewcommand{\thetable}{4}
\caption{QFT circuit benchmarks on Zeus-256.}\label{tab:4}
\begin{tabular}{@{}L{0.340000\tablewidth}R{0.075000\tablewidth}R{0.105000\tablewidth}R{0.250000\tablewidth}R{0.230000\tablewidth}@{}}
\toprule
\textbf{Problem} & \textbf{Qubits} & \textbf{QASM gates} & \textbf{Hardware execution wall time (ms; 10 shots)} & \textbf{Fluorescence interpretation (ms)} \\
\midrule
qft\_n12 & 12 & 84 & 0.1626 & 0.1772 \\
qft\_n13 & 13 & 97 & 0.1621 & 0.1768 \\
qft\_n14 & 14 & 112 & 0.1657 & 0.1804 \\
qft\_n17 & 17 & 161 & 0.2217 & 0.1937 \\
qft\_n18 & 18 & 180 & 0.2194 & 0.1946 \\
qft\_n19 & 19 & 199 & 0.2245 & 0.1968 \\
qft\_n63 & 63 & 2,047 & 1.4783 & 0.2739 \\
qft\_n64 & 64 & 2,112 & 0.8681 & 0.2957 \\
qft\_n127 & 127 & 8,191 & 1.4631 & 0.3831 \\
qft\_n128 & 128 & 8,320 & 1.5247 & 0.3872 \\
qft\_n255 & 255 & 32,767 & 3.1459 & 0.6068 \\
qft\_n256 & 256 & 33,024 & 2.9960 & 0.6055 \\
\bottomrule
\end{tabular}
\end{table}

\FloatBarrier

\subsection{Arithmetic and miscellaneous QASM benchmarks}\label{sec:6.5}

The general QASM benchmark suite was selected to span both width and algorithmic depth. It includes arithmetic adders and multipliers, Bernstein--Vazirani, satisfiability, Ising, basis-trotter, VQE-UCCSD, and other representative circuits. The largest circuit by original QASM gate count in this set was multiplier\_w42\_n254, with 254 processor-level logical qubits and 172,960 QASM gates. Its 10-shot hardware execution wall time was 96,213.4 ms. By contrast, the 256-qubit adder\_w127\_n256 circuit contained 4,450 QASM gates and had a hardware execution wall time of 66.9155 ms, illustrating the strong dependence of processor wall time on circuit structure and transpiled pulse workload rather than on qubit count alone.

\begin{table}[htbp]
\centering
\small
\setlength{\tabcolsep}{3pt}
\setlength{\tablewidth}{\dimexpr\linewidth-8\tabcolsep\relax}
\renewcommand{\arraystretch}{1.18}
\renewcommand{\thetable}{5}
\caption{Miscellaneous and arithmetic QASM circuit benchmarks on Zeus-256. A dash (---) indicates that a fluorescence-data interpretation timing value was not available in the benchmark record.}\label{tab:5}
\begin{tabular}{@{}L{0.340000\tablewidth}R{0.075000\tablewidth}R{0.105000\tablewidth}R{0.250000\tablewidth}R{0.230000\tablewidth}@{}}
\toprule
\textbf{Problem} & \textbf{Qubits} & \textbf{QASM gates} & \textbf{Hardware execution wall time (ms; 10 shots)} & \textbf{Fluorescence interpretation (ms)} \\
\midrule
cc\_n18 & 18 & 109 & 3.0303 & --- \\
multiplier\_n15 & 15 & 70 & 1.2921 & 0.0293 \\
seca\_n11 & 11 & 72 & 2.0954 & --- \\
bv\_n19 & 19 & 56 & 1.6921 & 0.0441 \\
bigadder\_n18 & 18 & 60 & 1.1148 & 0.0368 \\
sat\_n11 & 11 & 91 & 3.3860 & 0.0340 \\
ising\_n10 & 10 & 480 & 161.8960 & 0.0548 \\
basis\_trotter\_n4 & 4 & 1,506 & 2.2064 & 0.0163 \\
vqe\_uccsd\_n6 & 6 & 2,282 & 39.7510 & 0.0264 \\
vqe\_uccsd\_n8 & 8 & 10,808 & 891.0140 & 0.0385 \\
adder\_w4\_n10 & 10 & 145 & 2.4238 & 0.0393 \\
adder\_w8\_n18 & 18 & 285 & 2.4551 & 0.0607 \\
adder\_w16\_n34 & 34 & 565 & 3.6729 & 0.0852 \\
adder\_w32\_n66 & 66 & 1,125 & 7.3284 & 0.1397 \\
adder\_w64\_n130 & 130 & 2,245 & 20.8005 & 0.2477 \\
adder\_w127\_n256 & 256 & 4,450 & 66.9155 & 0.4529 \\
multiplier\_w2\_n14 & 14 & 398 & 3.7790 & 0.0408 \\
multiplier\_w4\_n26 & 26 & 1,580 & 61.8332 & 0.0830 \\
multiplier\_w8\_n50 & 50 & 6,292 & 445.8960 & 0.1211 \\
multiplier\_w16\_n98 & 98 & 25,124 & 3,929.0900 & 0.2178 \\
multiplier\_w32\_n194 & 194 & 100,420 & 36,997.5000 & 0.4046 \\
multiplier\_w42\_n254 & 254 & 172,960 & 96,213.4000 & 0.5632 \\
\bottomrule
\end{tabular}
\end{table}

\FloatBarrier

\subsection{Shor-family factorization staircase to 256 qubits}\label{sec:6.6}

We completed a Shor-family factorization staircase on Zeus-256 using a Gidney--Ekerå-inspired Ekerå--Håstad implementation with windowed exact arithmetic \cite{ref19,ref24,ref26}. The benchmark covers 11 test integers of 8, 16, 24, 32, 40, 48, 56, 64, 72, 80, and 84 bits. The experimental summary records the exact integer N, qubit count, gate count, number of attempts, execution time, and factor-recovery flag for each instance. All 11 instances report factor recovery, with 10 attempts recorded for each. The complete QASM circuit files for all 11 instances are provided as supplementary material at \url{https://github.com/DynexQuantum/zeus_paper}.

Table~\ref{tab:6a} retains the Gidney--Ekerå abstract resource model \cite{ref24} as a planning reference. Its estimates are \(Q_{\mathrm{GE}}(n)=3n+0.002n\log_2(n)\) logical qubits, \(T_{\mathrm{GE}}(n)=0.3n^3+0.0005n^3\log_2(n)\) Toffoli gates, and \(D_M(n)=500n^2+n^2\log_2(n)\) measurement depth, excluding distillation, routing, and error-correction overheads. Qubit estimates are rounded upward; Toffoli and depth values are rounded to the nearest integer. These estimates are distinct from the implementation resources reported in Table~\ref{tab:6b}. The observed register widths follow \(Q(n)=3n+4\) at every tested size: 28 qubits at 8 bits and 256 qubits at 84 bits, compared with the reference estimate of 254 qubits at 84 bits. The reference and Zeus processor-level qubit counts are compared only as register widths, without asserting equivalence of their error-correction encodings or physical overheads.

\begin{table}[htbp]
\centering
\small
\setlength{\tabcolsep}{3pt}
\setlength{\tablewidth}{\dimexpr\linewidth-6\tabcolsep\relax}
\renewcommand{\arraystretch}{1.18}
\renewcommand{\thetable}{6a}
\caption{Gidney--Ekerå abstract resource estimates retained as a planning reference \cite{ref24}. These are model estimates, not measured Zeus circuit resources; the completed implementation counts are reported in Table~\ref{tab:6b}.}\label{tab:6a}
\begin{tabular}{@{}L{0.190000\tablewidth}R{0.260000\tablewidth}R{0.260000\tablewidth}R{0.290000\tablewidth}@{}}
\toprule
\textbf{Integer size n (bits)} & \textbf{G--E logical qubits \(Q_{\mathrm{GE}}\) (ceil)} & \textbf{G--E Toffoli estimate \(T_{\mathrm{GE}}\)} & \textbf{G--E measurement depth \(D_M\)} \\
\midrule
8 & 25 & 154 & 32,192 \\
16 & 49 & 1,237 & 129,024 \\
24 & 73 & 4,179 & 290,641 \\
32 & 97 & 9,912 & 517,120 \\
40 & 121 & 19,370 & 808,515 \\
48 & 145 & 33,486 & 1,164,868 \\
56 & 169 & 53,195 & 1,586,212 \\
64 & 193 & 79,430 & 2,072,576 \\
72 & 217 & 113,126 & 2,623,985 \\
80 & 242 & 155,218 & 3,240,460 \\
84 & 254 & 179,706 & 3,573,104 \\
\bottomrule
\end{tabular}
\end{table}

Table~\ref{tab:6b} gives the completed measurements. The register widths span 28--256 processor-level logical qubits and the reported gate counts span 27,274--40,250,342. The gate column is reproduced from the experimental summary and is kept distinct from the theoretical Toffoli counts in Table~\ref{tab:6a}; the summary does not specify a gate set or decomposition level. Execution times are retained under the source label and at its three-decimal precision. The binary factor-recovery flag is an instance-level outcome, whereas the attempts column records the number of attempts for that instance.

\begin{table}[htbp]
\centering
\small
\setlength{\tabcolsep}{3pt}
\setlength{\tablewidth}{\dimexpr\linewidth-12\tabcolsep\relax}
\renewcommand{\arraystretch}{1.18}
\renewcommand{\thetable}{6b}
\caption{Completed Zeus-256 Shor-family factorization staircase using Gidney--Ekerå-inspired Ekerå--Håstad circuits with windowed exact arithmetic. Execution time is reproduced as reported; 10 attempts are recorded per instance.}\label{tab:6b}
\begin{tabular}{@{}C{0.088889\tablewidth}R{0.333333\tablewidth}R{0.077778\tablewidth}R{0.161111\tablewidth}C{0.088889\tablewidth}R{0.161111\tablewidth}C{0.088889\tablewidth}@{}}
\toprule
\textbf{Integer bits} & \textbf{Test integer N} & \textbf{Qubits} & \textbf{Gates (reported)} & \textbf{Attempts} & \textbf{Execution time (ms)} & \textbf{Factor found} \\
\midrule
8 & 143 & 28 & 27,274 & 10 & 188.574 & yes \\
16 & 49447 & 52 & 165,623 & 10 & 239.307 & yes \\
24 & 13423807 & 76 & 670,739 & 10 & 553.336 & yes \\
32 & 3108758101 & 100 & 1,690,447 & 10 & 552.622 & yes \\
40 & 827818182343 & 124 & 4,951,487 & 10 & 502.102 & yes \\
48 & 221674147038989 & 148 & 8,242,823 & 10 & 553.984 & yes \\
56 & 51096747444892003 & 172 & 12,690,289 & 10 & 554.955 & yes \\
64 & 14979923942189324171 & 196 & 18,625,853 & 10 & 549.343 & yes \\
72 & 3223144761104755519061 & 220 & 26,037,497 & 10 & 549.582 & yes \\
80 & 1032351713483218113671551 & 244 & 35,245,243 & 10 & 552.226 & yes \\
84 & 16503807113155763618449753 & 256 & 40,250,342 & 10 & 551.694 & yes \\
\bottomrule
\end{tabular}
\end{table}

Factor recovery was reported for all 11 tested integers across 110 recorded attempts in total. The largest successful instance was N = 16503807113155763618449753 (84 bits), using all 256 processor-level logical qubits and 40,250,342 reported gates, with an execution time of 551.694 ms. Across the staircase, the reported execution times range from 188.574 to 554.955 ms and are not monotonic in integer size or gate count. These values characterize the supplied runs and do not establish a timing scaling law.

The summary does not include recovered divisors, per-attempt measurement outcomes or success counts, timing dispersion, or a breakdown of readout and classical post-processing. Accordingly, the 11 positive instance-level flags are not interpreted as 110 successful attempts, and no per-attempt success probability or confidence interval is inferred. The timing aggregation and boundaries remain unspecified, as described in Section~\ref{sec:5.8}.

\FloatBarrier

\subsection{Application, variational, neural-network, and graph-state benchmarks}\label{sec:6.7}

The remaining benchmark families test structurally different workloads rather than a single scaling law. Portfolio-VQE circuits were executed at 16--18 processor-level logical qubits; travelling-salesman circuits at 9 and 16 processor-level logical qubits; routing circuits at 6 and 12 processor-level logical qubits; quantum neural network circuits at 17--21 processor-level logical qubits; compact VQE circuits at 12--16 processor-level logical qubits; and graph-state circuits at 16--20 processor-level logical qubits.

\begin{table}[htbp]
\centering
\small
\setlength{\tabcolsep}{3pt}
\setlength{\tablewidth}{\dimexpr\linewidth-8\tabcolsep\relax}
\renewcommand{\arraystretch}{1.18}
\renewcommand{\thetable}{7}
\caption{Portfolio-VQE circuit benchmarks on Zeus-256.}\label{tab:7}
\begin{tabular}{@{}L{0.340000\tablewidth}R{0.075000\tablewidth}R{0.105000\tablewidth}R{0.250000\tablewidth}R{0.230000\tablewidth}@{}}
\toprule
\textbf{Problem} & \textbf{Qubits} & \textbf{QASM gates} & \textbf{Hardware execution wall time (ms; 10 shots)} & \textbf{Fluorescence interpretation (ms)} \\
\midrule
portfolio\_vqe\_n16 & 16 & 424 & 645.8330 & 0.0791 \\
portfolio\_vqe\_n17 & 17 & 476 & 1,255.3300 & 0.0833 \\
portfolio\_vqe\_n18 & 18 & 531 & 3,003.6300 & 0.0895 \\
\bottomrule
\end{tabular}
\end{table}

\begin{table}[htbp]
\centering
\small
\setlength{\tabcolsep}{3pt}
\setlength{\tablewidth}{\dimexpr\linewidth-8\tabcolsep\relax}
\renewcommand{\arraystretch}{1.18}
\renewcommand{\thetable}{8}
\caption{Travelling-salesman circuit benchmarks on Zeus-256.}\label{tab:8}
\begin{tabular}{@{}L{0.340000\tablewidth}R{0.075000\tablewidth}R{0.105000\tablewidth}R{0.250000\tablewidth}R{0.230000\tablewidth}@{}}
\toprule
\textbf{Problem} & \textbf{Qubits} & \textbf{QASM gates} & \textbf{Hardware execution wall time (ms; 10 shots)} & \textbf{Fluorescence interpretation (ms)} \\
\midrule
tsp\_n16 & 16 & 171 & 437.1860 & 0.0760 \\
tsp\_n9 & 9 & 94 & 38.3354 & 0.0358 \\
\bottomrule
\end{tabular}
\end{table}

\begin{table}[htbp]
\centering
\small
\setlength{\tabcolsep}{3pt}
\setlength{\tablewidth}{\dimexpr\linewidth-8\tabcolsep\relax}
\renewcommand{\arraystretch}{1.18}
\renewcommand{\thetable}{9}
\caption{Routing circuit benchmarks on Zeus-256.}\label{tab:9}
\begin{tabular}{@{}L{0.340000\tablewidth}R{0.075000\tablewidth}R{0.105000\tablewidth}R{0.250000\tablewidth}R{0.230000\tablewidth}@{}}
\toprule
\textbf{Problem} & \textbf{Qubits} & \textbf{QASM gates} & \textbf{Hardware execution wall time (ms; 10 shots)} & \textbf{Fluorescence interpretation (ms)} \\
\midrule
routing\_n12 & 12 & 81 & 27.1287 & 0.0339 \\
routing\_n6 & 6 & 39 & 3.2677 & 0.0201 \\
\bottomrule
\end{tabular}
\end{table}

\begin{table}[htbp]
\centering
\small
\setlength{\tabcolsep}{3pt}
\setlength{\tablewidth}{\dimexpr\linewidth-8\tabcolsep\relax}
\renewcommand{\arraystretch}{1.18}
\renewcommand{\thetable}{10}
\caption{Quantum neural network circuit benchmarks on Zeus-256.}\label{tab:10}
\begin{tabular}{@{}L{0.340000\tablewidth}R{0.075000\tablewidth}R{0.105000\tablewidth}R{0.250000\tablewidth}R{0.230000\tablewidth}@{}}
\toprule
\textbf{Problem} & \textbf{Qubits} & \textbf{QASM gates} & \textbf{Hardware execution wall time (ms; 10 shots)} & \textbf{Fluorescence interpretation (ms)} \\
\midrule
dnn\_n17 & 17 & 934 & 2,382.3100 & 0.1268 \\
dnn\_n19 & 19 & 1,158 & 9,690.3200 & 0.0565 \\
dnn\_n21 & 21 & 1,406 & 56,875.0000 & 0.0620 \\
\bottomrule
\end{tabular}
\end{table}

\begin{table}[htbp]
\centering
\small
\setlength{\tabcolsep}{3pt}
\setlength{\tablewidth}{\dimexpr\linewidth-8\tabcolsep\relax}
\renewcommand{\arraystretch}{1.18}
\renewcommand{\thetable}{11}
\caption{VQE circuit benchmarks on Zeus-256.}\label{tab:11}
\begin{tabular}{@{}L{0.340000\tablewidth}R{0.075000\tablewidth}R{0.105000\tablewidth}R{0.250000\tablewidth}R{0.230000\tablewidth}@{}}
\toprule
\textbf{Problem} & \textbf{Qubits} & \textbf{QASM gates} & \textbf{Hardware execution wall time (ms; 10 shots)} & \textbf{Fluorescence interpretation (ms)} \\
\midrule
vqe\_n12 & 12 & 58 & 40.4191 & 0.0306 \\
vqe\_n14 & 14 & 68 & 67.8201 & 0.0330 \\
vqe\_n16 & 16 & 78 & 423.9670 & 0.0486 \\
\bottomrule
\end{tabular}
\end{table}

\begin{table}[htbp]
\centering
\small
\setlength{\tabcolsep}{3pt}
\setlength{\tablewidth}{\dimexpr\linewidth-8\tabcolsep\relax}
\renewcommand{\arraystretch}{1.18}
\renewcommand{\thetable}{12}
\caption{Graph-state circuit benchmarks on Zeus-256.}\label{tab:12}
\begin{tabular}{@{}L{0.340000\tablewidth}R{0.075000\tablewidth}R{0.105000\tablewidth}R{0.250000\tablewidth}R{0.230000\tablewidth}@{}}
\toprule
\textbf{Problem} & \textbf{Qubits} & \textbf{QASM gates} & \textbf{Hardware execution wall time (ms; 10 shots)} & \textbf{Fluorescence interpretation (ms)} \\
\midrule
graph\_state\_n16 & 16 & 32 & 1.4156 & 0.0457 \\
graph\_state\_n18 & 18 & 36 & 1.5114 & 0.0500 \\
graph\_state\_n20 & 20 & 40 & 1.9465 & 0.0526 \\
\bottomrule
\end{tabular}
\end{table}

\FloatBarrier

\subsection{Grover search experiment}\label{sec:6.8}

The Grover experiment \cite{ref15} used the supplied OpenQASM 2.0 three-qubit circuit with marked output bitstring 101. The circuit first prepares the uniform superposition using H(q[0]), H(q[1]), and H(q[2]). Each oracle marks the state q[2]q[1]q[0] = 101 by temporarily inverting q[1] and implementing a phase-marking construction with H(q[2]), CCX(q[0],q[1],q[2]), and H(q[2]), followed by restoration of q[1]. The oracle is followed by the standard three-qubit diffusion sequence. This oracle--diffusion pair is applied twice. In total, the QASM circuit contains 43 unitary gates---23 H, 16 X, and four CCX gates---followed by three measurements. The exact supplied three-qubit circuit is shown in Figure~\ref{fig:8}.

The circuit was executed for 1,024 measurement shots. The processor returned the target state 101 in 958 shots, corresponding to a measured success probability of 93.55\%. The 95\% Wilson confidence interval was 91.88\%--94.90\%. An independently computed ideal circuit reference yielded 94.53\% target-state probability, while uniform random guessing over the eight computational-basis states would yield 12.5\%. The complete measured counts for bitstrings 000, 001, 010, 011, 100, 101, 110, and 111 were 8, 5, 17, 6, 11, 958, 7, and 12, respectively. The observed distribution is therefore strongly concentrated on the marked state and closely tracks the independent ideal reference, providing an experimental demonstration of amplitude amplification on Zeus. This experiment is used as an algorithm-level validation of the processor and is not presented as evidence of computational speedup over classical search.

\begin{figure}[htbp]
\centering
\renewcommand{\thefigure}{6}\renewcommand{\theHfigure}{6}
\includegraphics[width=\linewidth,height=0.69\textheight,keepaspectratio]{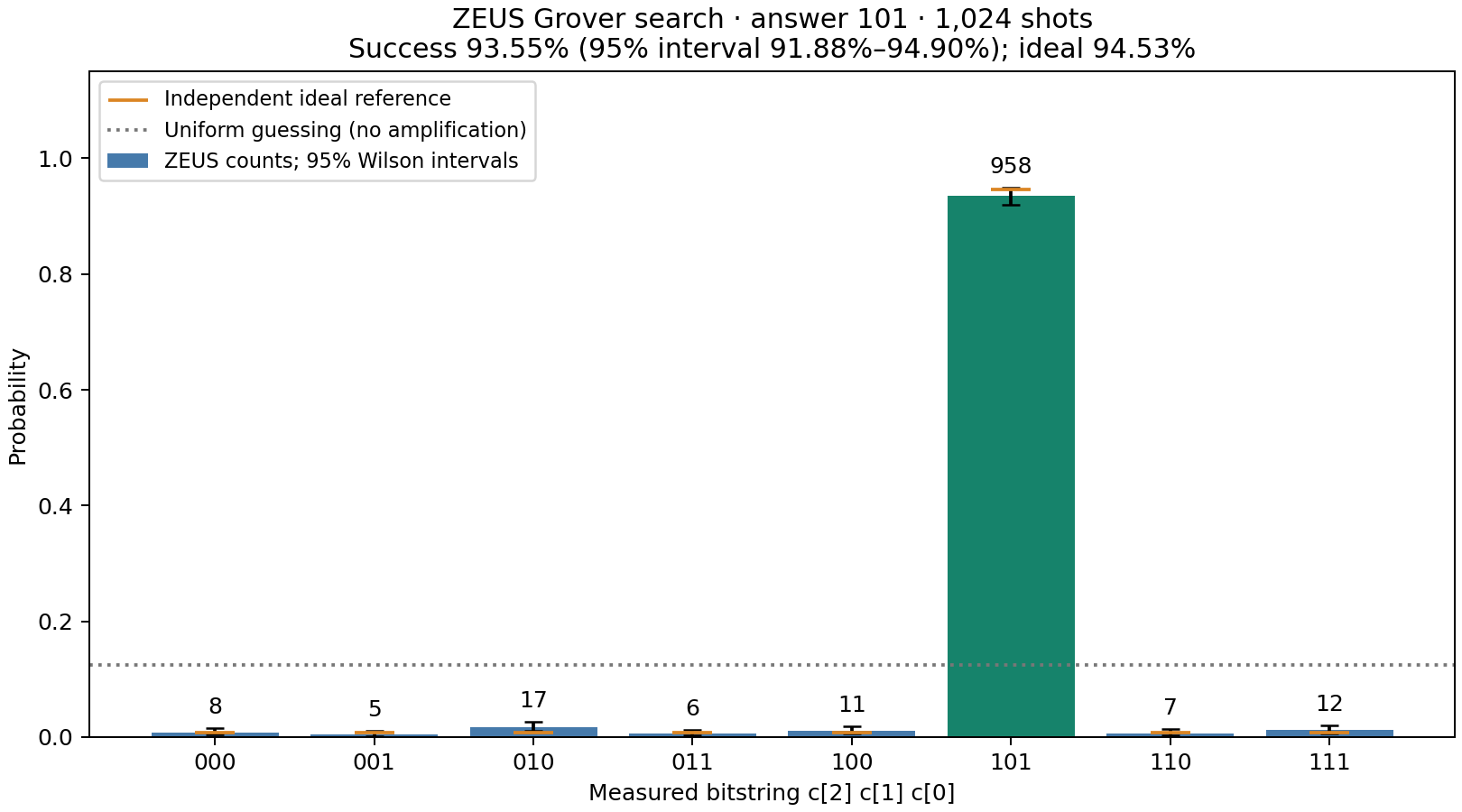}
\caption{Grover-search output distribution on Zeus for target bitstring 101 using 1,024 shots. The measured success probability was 93.55\% (95\% Wilson interval 91.88\%--94.90\%), compared with an independent ideal reference of 94.53\% and a uniform-guessing baseline of 12.5\%.}\label{fig:6}
\end{figure}

\FloatBarrier

\subsection{Representative circuit structures}\label{sec:6.9}

\begin{figure}[htbp]
\centering
\renewcommand{\thefigure}{7}\renewcommand{\theHfigure}{7}
\includegraphics[width=\linewidth,height=0.69\textheight,keepaspectratio]{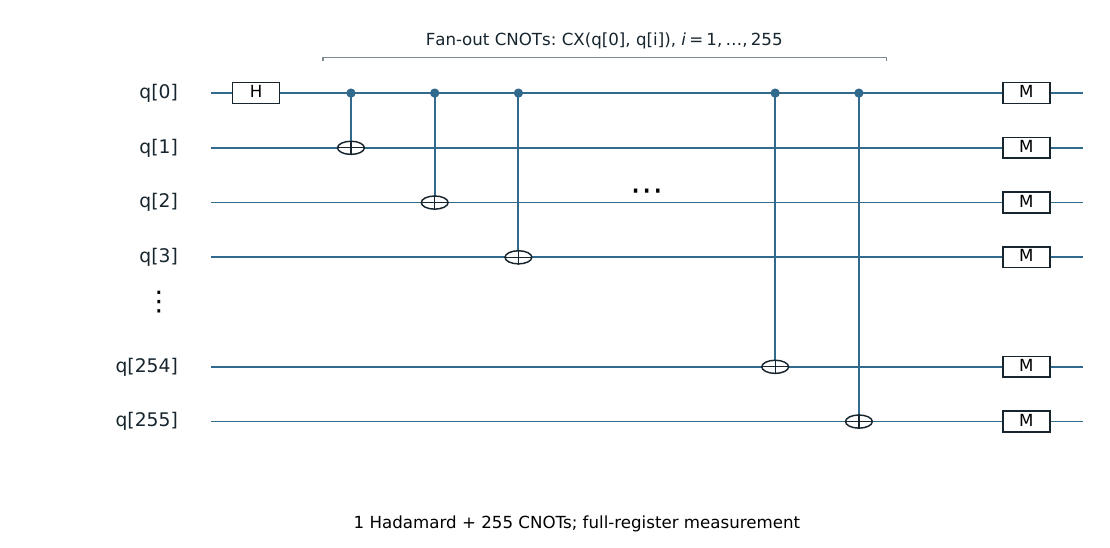}
\caption{Schematic circuit representation of the 256-qubit GHZ-state preparation used in the Zeus-256 validation experiment. The OpenQASM 2.0 circuit applies H to q[0], followed by 255 fan-out controlled-NOT operations CX(q[0],q[i]) for \(i=1,\ldots,255\), and concludes with full-register measurement. The ideal pre-measurement target state is \(|\mathrm{GHZ}_{256}\rangle=\frac{|0\rangle^{\otimes256}+|1\rangle^{\otimes256}}{\sqrt{2}}\).}\label{fig:7}
\end{figure}

\begin{figure}[htbp]
\centering
\renewcommand{\thefigure}{8}\renewcommand{\theHfigure}{8}
\includegraphics[width=\linewidth,height=0.69\textheight,keepaspectratio]{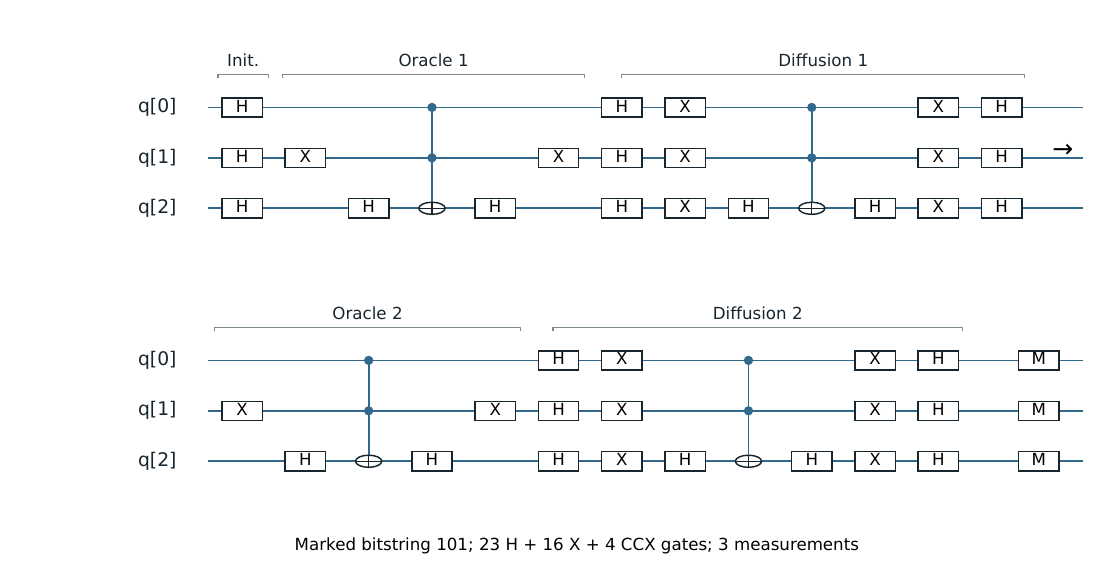}
\caption{Circuit representation of the supplied three-qubit Grover-search experiment targeting output bitstring 101. The circuit prepares a uniform superposition, applies two oracle--diffusion iterations, and then measures all three qubits. Excluding the final measurements, the supplied QASM circuit contains 43 unitary gates: 23 Hadamard gates, 16 X gates, and four Toffoli (CCX) gates.}\label{fig:8}
\end{figure}

\begin{figure}[htbp]
\centering
\renewcommand{\thefigure}{9}\renewcommand{\theHfigure}{9}
\includegraphics[width=\linewidth,height=0.69\textheight,keepaspectratio]{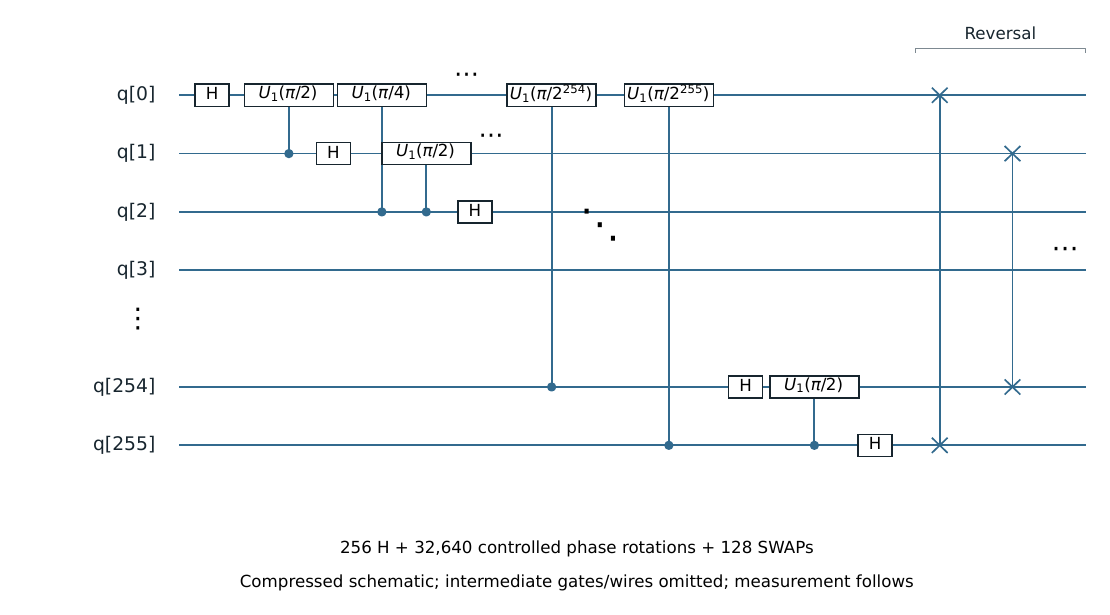}
\caption{Schematic circuit representation of the 256-qubit coherent quantum Fourier transform benchmark. The OpenQASM 2.0 circuit applies a Hadamard gate to each qubit, followed by a triangular network of controlled-U1 phase rotations with \(\theta_{j,k}=\pi/2^{j-k}\) for j > k, and concludes with 128 SWAP gates that reverse the qubit order. The full-register experiment used a randomly assigned input state rather than a fixed all-zero input. Excluding the final register-wide measurement statement, the circuit contains 33,024 QASM gates: 256 Hadamard gates, 32,640 controlled-U1 phase rotations, and 128 SWAP gates.}\label{fig:9}
\end{figure}

Figures~\ref{fig:7}--\ref{fig:9} provide representative circuit drawings for the three experimentally highlighted circuits. For the 256-qubit GHZ and QFT experiments, the drawings use a compressed schematic representation with omitted intermediate wires or gates indicated by ellipses; the underlying QASM descriptions remain the authoritative circuit definitions.

\FloatBarrier

\subsection{Timing interpretation and cross-circuit observations}\label{sec:6.10}

All timing values in the completed benchmark Tables \ref{tab:3}--\ref{tab:5} and \ref{tab:7}--\ref{tab:12} use 10 shots per benchmark entry. The completed Shor measurements in Table~\ref{tab:6b} record 10 factorization attempts per instance and a separately reported execution-time field whose aggregation and timing boundaries are unspecified. They are therefore described in Section~\ref{sec:6.6} separately from this 10-shot timing comparison. Hardware execution wall time, \(T_{\mathrm{hw}}\), denotes the recorded post-transpilation processor-control execution time and excludes physical fluorescence photon acquisition and subsequent fluorescence-data interpretation. The second timing quantity, \(T_{\mathrm{interpretation}}\), is fluorescence-data interpretation time and likewise excludes photon acquisition. The tables therefore do not represent complete end-to-end application latency. Values are reproduced at the precision returned by the control/acquisition stack, and no formal uncertainty is implied by the final displayed decimal place.

The timing observations show that qubit count alone is not sufficient to explain hardware execution wall time. QFT spans 12 to 256 processor-level logical qubits and 84 to 33,024 QASM gates while remaining in the millisecond regime for the reported 10-shot wall-time metric. The QFT width series is not monotonic---for example, the recorded 64-qubit value is lower than the 63-qubit value---consistent with the fact that compilation, scheduling, pulse fusion, and circuit structure can alter the resulting hardware workload. The GHZ series ranges from 21 to 256 processor-level logical qubits, while deep arithmetic and application circuits at substantially smaller widths can require orders of magnitude more wall time. The 254-qubit, 172,960-gate multiplier required 96.2134 s for 10 shots, while a 21-qubit QNN circuit required 56.875 s. Because run-to-run variance is not reported, these data are interpreted as descriptive cross-circuit timing observations rather than as a smooth or fitted scaling law.

\FloatBarrier

\section{Discussion}\label{sec:7}

The experiments reported here reposition the architecture from a predominantly conceptual description to an experimentally benchmarked processor with 256 processor-level logical qubits. Zeus-256 operates a 16 \(\times\) 16 register of simultaneously active processor-level logical qubits at room temperature and combines electron-spin ensemble computation, nuclear-spin memory and ancillary resources, MW/RF pulse control, optical initialization and fluorescence readout, programmable biomolecular quantum-bus coupling, and measurement-conditioned control within a common execution stack. This experimental register is distinct from the 9,984-qubit scale target discussed in Section~\ref{sec:3.8}.

The Bell--CHSH measurements \cite{ref17,ref18} provide an independent two-qubit test of nonclassical correlations. With 1,024 shots per setting, the measured CHSH parameter was S = 2.8008 \(\pm\) 0.0446, while the bootstrap analysis over 20,000 resamples yielded mean S = 2.80131 and \(\sigma\) = 0.04442. The result is substantially above the local-classical bound S = 2 and approaches the Tsirelson bound \(2\sqrt{2}\). This experiment complements the larger GHZ benchmark family: the Bell experiment establishes a directly interpretable two-qubit nonclassical correlation test, while the GHZ experiments exercise the processor's multipartite state-preparation capabilities over register sizes extending to 256 processor-level logical qubits. In Zeus-256, the entangling interaction resource for such compiled circuits is the proprietary biomolecular quantum-bus layer. In the 256-qubit GHZ fan-out circuit, interactions between q[0] and spatially separated processor fields are mapped to bus-mediated pathways, so the experiment provides processor-level evidence for long-range bus-enabled entangling connectivity without disclosing the microscopic bus implementation.

The benchmark suite also shows that hardware execution wall time cannot be inferred from qubit count alone. Full-register QFT and GHZ circuits remain in the millisecond regime for the reported 10-shot wall-time metric, whereas deep arithmetic, QNN, VQE-UCCSD, and application circuits at smaller widths can require orders of magnitude more wall time. The Grover experiment provides a complementary algorithm-level validation: for the marked state 101, Zeus produced 958 successful outcomes from 1,024 shots, corresponding to 93.55\% success, close to the independent ideal reference of 94.53\% and far above the 12.5\% uniform-guessing baseline. Consistent with Grover's amplitude-amplification mechanism \cite{ref15}, this result demonstrates amplitude amplification on the processor but is not interpreted as evidence of end-to-end computational speedup. Overall, the observed timing behavior is consistent with the pulse-native execution model described in Section~\ref{sec:5}: original QASM gate count is only an algorithmic descriptor, while the actual hardware workload is determined by the transpiled MW/RF pulse schedule, coupling requirements, pulse timing, and circuit structure.

The programmable biomolecular quantum-bus layer is central to the connectivity model. It is not treated merely as a passive dielectric: the proprietary material uses a protein--chromophore architecture similar to EYFP and incorporates a metastable electronic triplet spin degree of freedom as an internal molecular mediator. Fluorescent-protein chromophores such as EYFP have independently been demonstrated as optically addressable spin qubits \cite{ref25}, providing an external precedent for this class of biomolecular quantum degree of freedom. In Zeus-256, however, the molecular spin subsystem is not a user-exposed processor qubit; it forms part of the shared quantum-bus coupling fabric. Optical programming changes the material's dielectric and coupling state, reshaping the MW/RF environment and establishing calibrated bus-mediated interaction pathways between selected fields. Because the bus can connect spatially separated processor fields, nonlocal circuit edges need not be implemented exclusively through intermediate nearest-neighbour SWAP chains. The exact molecular identity, internal triplet-state dynamics, programming maps, bus-transfer characteristics, native entangling pulse sequence, and quantitative coupling-characterization data are proprietary and are not disclosed. The scientific evidence reported here is therefore processor-level: the integrated bus and control stack is evaluated through compiled long-range multi-qubit interactions, Bell correlations, multipartite GHZ-state preparation, and algorithm-level measurements.

The results should also be interpreted in the context of prior room-temperature NV-diamond work. Earlier experiments established coherent electron--nuclear registers, multipartite entanglement, quantum error correction, and logical operations in substantially smaller local registers \cite{ref1,ref2,ref3,ref5,ref6,ref7}. Reconfigurable or highly connected processors have also been demonstrated in other modalities, including neutral atoms and trapped ions, using different physical mechanisms \cite{ref10,ref11}. Zeus-256 addresses a different systems question: whether a room-temperature NV-diamond register can integrate 256 simultaneously active processor-level logical qubits with software-defined pulse execution and optically programmable biomolecular quantum-bus connectivity.

The 9,984-qubit configuration is intentionally treated as an engineering scale-up of the demonstrated architecture rather than as a separate research concept. The qubit modality, processor-level logical-qubit construction, optical initialization/readout principle, MW/RF control model, biomolecular quantum-bus coupling principle, and compiler-to-pulse execution path are already instantiated at Zeus-256 scale. Scaling the same architecture to 9,984 processor-level logical qubits would primarily require processor-field replication and densification, control-channel expansion, RFSoC bandwidth scaling, packaging, calibration automation, optical-uniformity engineering, bus-configuration/calibration throughput, crosstalk management, manufacturing yield improvement, and system integration. On the basis of the demonstrated architecture, this transition is not presently expected to require a different qubit modality or coupling principle; validation at increased scale remains necessary.

The present benchmark data establish processor operation and provide cross-circuit timing observations across a broad set of quantum-circuit structures, but they do not by themselves establish quantum computational advantage over classical computing. Any such claim would require problem-matched classical baselines, equivalent solution-quality criteria, and end-to-end resource accounting. The current work therefore restricts its conclusions to experimentally demonstrated processor operation, quantum-state validation, integrated programmable-connectivity functionality, and circuit-level execution characteristics.

\FloatBarrier

\section{Limitations}\label{sec:8}

Several limitations should be considered when interpreting the results. First, the term processor-level logical qubit in Zeus-256 refers to the processor variable formed from an electron-spin ensemble together with nuclear-spin memory and ancillary resources. It is an implementation-level term and is not intended to imply the experimentally demonstrated fault-tolerant logical error suppression associated with encoded logical qubits in the stricter quantum-error-correction sense. The internal control and error-management protocol is proprietary. A complete scientific characterization of the logical layer would require observable before/after error-management metrics or another quantitative logical-error measure in addition to the architectural description.

Second, the 256-qubit GHZ validation uses population and parity measurements rather than full density-matrix tomography \cite{ref20,ref21}. The complete 30,000-shot record and acquisition summary accompany the manuscript as Supplementary Data 1 and Supplementary Note 1. Fourier projection of 257 rotated-basis settings yields the target-phase coherence A = 0.997920058 and \(F_{\mathrm{GHZ}}=(P+A)/2=0.998960029\), with estimated SE = 0.002885024 and conservative 95\% Hoeffding interval [0.982341653, 1] after intersection with the physical range. The entanglement interpretation assumes trusted local basis rotations and readout, stable state preparation across settings, and independent trials. No additional readout/SPAM correction is applied in the count analysis. Upstream mitigation, rejected-shot information, and phase-calibration records are not documented in the supplied files, and systematic uncertainty from these sources is not included in the statistical interval.

Third, the benchmark timing quantities are intentionally partial. \(T_{\mathrm{hw}}\) is the recorded post-transpilation processor-control execution time and excludes circuit transpilation, physical fluorescence photon acquisition, subsequent fluorescence-data interpretation, queueing, and host-side communication. \(T_{\mathrm{interpretation}}\) reports fluorescence-data interpretation of already acquired data and likewise excludes the physical photon-acquisition interval. The tabulated values should therefore not be interpreted as complete end-to-end application latency. Run-to-run timing variance is not reported, so the width-series data are descriptive rather than a fitted scaling law. The Shor execution-time field in Table~\ref{tab:6b} has unspecified aggregation and timing boundaries and is not equated with \(T_{\mathrm{hw}}\). Its instance-level recovery flags do not provide per-attempt success probabilities; the supporting summary omits the recovered divisors and individual measurement records.

Fourth, the exact biomolecular material identity, microscopic quantum-bus dynamics, detailed fabrication process, optical programming procedure, calibrated interaction maps, bus-transfer characteristics, native inter-field entangling pulse implementation, and quantitative coupling-characterization data are proprietary and are not disclosed. The manuscript identifies the material only at the level necessary to describe the physical concept: a proprietary protein--chromophore system similar to EYFP with a metastable electronic triplet spin degree of freedom used internally as an optically addressable molecular mediator. Reference \cite{ref25} establishes EYFP as an external precedent for this class of fluorescent-protein spin qubit, but the proprietary Zeus material is not claimed to be EYFP. Reproducibility is consequently provided at the level of processor architecture, circuit definitions, measurement procedures, and externally observable processor-level results rather than at the level required to reproduce the proprietary biomolecular quantum bus itself.

Fifth, the experimental processor contains 256 simultaneously active processor-level logical qubits. The 9,984-qubit configuration is an engineering scale target and has not been demonstrated experimentally in this work. On the basis of reuse of the same demonstrated qubit modality, control, readout, coupling, and compilation principles, the transition is not presently expected to require a different qubit modality or coupling principle. Validation at increased scale is still required to confirm fabrication yield, calibration throughput, optical uniformity, MW/RF isolation, coupling behavior, bandwidth, thermal behavior, and aggregate processor-level state performance.

Finally, the heterogeneous benchmark suite is designed to characterize processor execution rather than to establish quantum advantage. No claim of classical intractability or superiority over the best classical algorithms is made from the timing tables alone.

\FloatBarrier

\section{Conclusion}\label{sec:9}

We experimentally characterize Zeus-256, a room-temperature NV-diamond quantum processor comprising 256 simultaneously active processor-level logical qubits arranged as a 16 \(\times\) 16 processor-field array. Each processor-level logical qubit combines an NV electron-spin ensemble used for computation with associated nuclear-spin memory and ancillary resources. OpenQASM 2.0 circuits are transpiled into RFSoC-controlled microwave, radio-frequency, optical, timing, measurement, and biomolecular-bus configuration sequences. An optically programmable biomolecular quantum-bus layer provides the reconfigurable coupling fabric. The proprietary material uses a protein--chromophore architecture similar to EYFP and incorporates a metastable electronic triplet spin degree of freedom as an internal molecular mediator. This subsystem is not exposed as an additional computational qubit register; it functions as part of the effective quantum bus enabling selected spatially separated processor fields to exchange quantum information and participate in mediated entangling operations.

The processor was benchmarked across QFT, GHZ entanglement, Grover search, Shor-family factorization, arithmetic and miscellaneous QASM, portfolio-VQE, travelling-salesman, routing, QNN, VQE, and graph-state circuits. The completed Shor staircase reports factor recovery for all 11 test integers spanning 8--84 bits, with 10 attempts recorded per instance. Its 84-bit endpoint, N = 16503807113155763618449753, uses all 256 processor-level logical qubits and 40,250,342 reported gates, with a reported execution time of 551.694 ms. These instance-level outcomes extend the completed benchmark set to factorization; per-attempt success probabilities and end-to-end factorization latency are not determined by the supplied summary. Independent Bell--CHSH measurements with 1,024 shots per setting produced S = 2.8008 \(\pm\) 0.0446, with a 20,000-resample bootstrap mean of 2.80131 and \(\sigma\) = 0.04442, providing direct evidence of nonclassical two-qubit correlations. A three-qubit Grover search targeting 101 produced 958 successful outcomes from 1,024 shots, corresponding to 93.55\% success (95\% Wilson interval 91.88\%--94.90\%) versus an independent ideal reference of 94.53\%. A separate full-register GHZ validation experiment used the 256-qubit OpenQASM circuit H(q[0]) followed by CX(q[0],q[i]) for \(i=1,\ldots,255\). Using 15,000 Z-basis shots and 15,000 shots distributed across 257 phase-rotated settings, population and parity measurements yielded a fixed-phase \(\mathrm{GHZ}^{+}\) fidelity of 99.8960\% with an estimated SE of 0.2885 percentage points and a conservative physical 95\% Hoeffding interval of [98.2342\%, 100\%]. The phase-resolved counts and acquisition summary accompany the manuscript.

The experimental results establish room-temperature processor operation, electron--nuclear-spin control, pulse-native circuit execution, optical initialization and fluorescence readout, and processor-level quantum-state validation at Zeus-256 scale. The demonstrated architecture also incorporates a proprietary biomolecular quantum bus with an internal protein--chromophore triplet-spin mediator similar in class to EYFP-based fluorescent-protein spin qubits \cite{ref25}. This molecular spin subsystem provides local and long-range programmable interaction pathways across the processor array but is not counted as an additional processor-level logical-qubit register. The full-register GHZ fan-out circuit exercises bus-mediated interactions between q[0] and processor fields distributed throughout the 16 \(\times\) 16 array, while Bell--CHSH measurements provide an independent two-qubit nonclassical-correlation benchmark. The exact molecular identity, microscopic triplet-state dynamics, coupling-transfer characteristics, and native entangling pulse implementation remain outside the disclosed experimental scope of this paper. On the basis of the demonstrated architecture, the proposed transition to 9,984 processor-level logical qubits is treated as an engineering scale-up and is not presently expected to require a different qubit modality or coupling principle.

\FloatBarrier

\section*{Data and Code Availability}

All supplementary files supporting this study are available at \url{https://github.com/DynexQuantum/zeus_paper}. The repository provides benchmark QASM circuits, qubit and gate counts, timing measurements, Bell--CHSH measurement outcomes and bootstrap analysis, Grover-search output counts and confidence-interval analysis, and the accompanying analysis and figure-generation files.

The complete GHZ-256 dataset is provided as Supplementary Data 1, \texttt{ghz\_counts\_fidelity.json}, containing qubit count, backend metadata, the Z-basis and all 257 rotated-basis count dictionaries, full-precision phase values, shots per setting, and the stored fidelity and uncertainty results. Supplementary Note 1, \texttt{GHZ-256.md}, provides the acquisition log and matching numerical summary. These filenames identify the publication versions of the two accompanying files. Section~\ref{sec:6.3.1} specifies the count-to-parity, Fourier, variance-propagation, and Hoeffding calculations required to reproduce the results. The JSON phase values, rather than the rounded phases in the log or tables, should be used for numerical reproduction.

The repository includes the complete QASM circuit files for all 11 Shor-family factorization instances in Section~\ref{sec:6.6}, spanning 8--84-bit integers and 28--256 processor-level logical qubits. The accompanying \texttt{table.md} contains the exact test integers, qubit and gate counts, attempt counts, execution-time values, and factor-recovery flags reproduced in Table~\ref{tab:6b}. The tabulated summary itself does not contain recovered divisors, individual measurement records, or detailed timing and post-processing logs.

The exact protein--chromophore identity and composition of the biomolecular quantum-bus material, microscopic triplet-state dynamics, optical programming procedures, calibrated coupling maps, coupling-characterization data, native inter-field entangling implementation, and logical-control/error-management implementation are proprietary and excluded from disclosure.

\FloatBarrier

\section*{Competing Interests}

Certain authors are inventors and/or applicants on patent applications covering aspects of the NV-diamond processor, programmable biomolecular quantum-bus coupling, measurement-conditioned and driven-dissipative control, and related quantum-processing technology.

\FloatBarrier

\end{document}